\documentclass[twocolumn]{aastex631}
\usepackage{amsmath}
\usepackage{CJK}

\newcommand\prospector{\texttt {Prospector}}
\newcommand\cloudy{\texttt{CLOUDY}}
\newcommand\cue{\texttt{Cue}}
\newcommand\QH{Q_\mathrm{H}}
\newcommand\nH{n_\mathrm{H}}

\newcommand\bX{\boldsymbol{X}}
\newcommand\btheta{\boldsymbol{\theta}}

\begin{document}
\begin{CJK*}{UTF8}{gbsn}

\title{Cue: A Fast and Flexible Photoionization Emulator for Modeling Nebular Emission Powered By Almost Any Ionizing Source}

\author[0000-0002-0682-3310]{Yijia Li (李轶佳)}
\email{yzl466@psu.edu}
\affiliation{Department of Astronomy \& Astrophysics, The Pennsylvania State University, University Park, PA 16802, USA}
\affiliation{Institute for Gravitation and the Cosmos, The Pennsylvania State University, University Park, PA 16802, USA}

\author[0000-0001-6755-1315]{Joel Leja}
\affiliation{Department of Astronomy \& Astrophysics, The Pennsylvania State University, University Park, PA 16802, USA}
\affiliation{Institute for Computational \& Data Sciences, The Pennsylvania State University, University Park, PA 16802, USA}
\affiliation{Institute for Gravitation and the Cosmos, The Pennsylvania State University, University Park, PA 16802, USA}

\author[0000-0002-9280-7594]{Benjamin D. Johnson}
\affiliation{Center for Astrophysics $\mid$ Harvard \& Smithsonian, Cambridge, MA, USA}

\author[0000-0002-8224-4505]{Sandro Tacchella}
\affiliation{Kavli Institute for Cosmology, University of Cambridge, Madingley Road, Cambridge, CB3 0HA, UK}
\affiliation{Cavendish Laboratory, University of Cambridge, 19 JJ Thomson Avenue, Cambridge, CB3 0HE, UK}

\author[0000-0002-3324-4824]{Rebecca Davies}
\affiliation{Centre for Astrophysics and Supercomputing, Swinburne University of Technology, Hawthorn, Victoria 3122, Australia}
\affiliation{ARC Centre of Excellence for All Sky Astrophysics in 3 Dimensions (ASTRO 3D), Australia}

\author[0000-0002-5615-6018]{Sirio Belli}
\affiliation{Dipartimento di Fisica e Astronomia, Universit{\`a} di Bologna, Bologna, Italy}

\author[0000-0002-8435-9402]{Minjung Park}
\affiliation{Center for Astrophysics $\mid$ Harvard \& Smithsonian, Cambridge, MA, USA}

\author[0000-0002-2791-5011]{Razieh \ Emami}
\affiliation{Center for Astrophysics $\vert$ Harvard \& Smithsonian, 60 Garden Street, Cambridge, MA 02138, USA}
%\author{Blue Jay team}

\begin{abstract}
The complex physics governing nebular emission in galaxies, particularly in the early universe, often defy simple low-dimensional models. This has proven to be a significant barrier in understanding the (often diverse) ionizing sources powering this emission. We present \cue\footnote{\cue\ v0.1 is available online at \url{https://github.com/yi-jia-li/cue}.}, a highly flexible tool for interpreting nebular emission across a wide range of abundances and ionizing conditions of galaxies at different redshifts. Unlike typical nebular models used to interpret extragalactic nebular emission, our model does not require a specific ionizing spectrum as a source, instead approximating the ionizing spectrum with a 4-part piece-wise power-law. We train a neural net emulator based on the \cloudy\ photoionization modeling code and make self-consistent nebular continuum and line emission predictions. Along with the flexible ionizing spectra, we allow freedom in [O/H], [N/O], [C/O], gas density, and total ionizing photon budget. This flexibility allows us to either marginalize over or directly measure the incident ionizing radiation, thereby directly interrogating the source of the ionizing photons in distant galaxies via their nebular emission. Our emulator demonstrates a high accuracy, with $\sim$1\% uncertainty in predicting the nebular continuum and $\sim$5\% uncertainty in the emission lines. Mock tests suggest \cue\ is well-calibrated and produces useful constraints on the ionizing spectra when $S/N (\mathrm{H}_\alpha) \gtrsim 10$, and furthermore capable of distinguishing between the ionizing spectra predicted by single and binary stellar models. The compute efficiency of neural networks facilitates future applications of \cue\ for rapid modeling of the nebular emission in large samples and Monte Carlo sampling techniques. 
%In the future, we plan to apply \cue\ on a variety of data, including proper marginalization over uncertainties when fitting broadband photometry and direct inference when fitting detailed spectroscopy.
\end{abstract}

\section{Introduction}
Nebular emission in galaxies probes the chemical abundances, density, and ionization states of the gas near sources of ionizing radiation. It plays a critical role in measuring the properties of these ionizing sources -- for example, measuring star formation rates (SFRs) or Active Galactic Nucleus (AGN) luminosities. With JWST opening a new window to observe galaxies in the reionization era, a long-standing challenge in observations of distant galaxies has a renewed importance: what sources are ionizing the gas in these galaxies, and how can we learn about their properties? Recent JWST observations have unveiled individual galaxies with unusual emission line properties in the early universe (e.g., \citealt{Arellano-Cordova2022}; \citealt{Brinchmann2023}; \citealt{Bunker2023}; \citealt{Cameron2023b}; \citealt{Curti2023a}; \citealt{Katz2023a}). These galaxies exhibit remarkable emission lines that challenge the nebular models calibrated by low-redshift normal star-forming galaxies. %To understand these galaxies and identify the sources of reionization, a general and flexible nebular emission tool is imperative. 
    
Moreover, the accuracy of nebular emission modeling has profound effects on the interpretation of the photometry and spectra of high-redshift galaxies, as nebular emission contributes $\gtrsim$20\% of UV and optical emission of star-forming galaxies (e.g., \citealt{Reines2010}; \citealt{Stark2013}; \citealt{Pacifici2015}), with increasing importance towards the high specific star formation rates (sSFRs) measured at high redshifts. The uncertainties associated with the nebular model will not only affect the inferred ionizing gas properties but also influence all the estimated galaxy properties including mass, SFR, dust properties, etc. 
%Previous nebular models based on galaxies at $z<3.5$ rely on assumptions that remain largely unconstrained at high redshift. 
While previous nebular emission studies are often based on galaxies at $z<3.5$, high-redshift galaxies exhibit different nebular conditions and ionizing sources, such as a high ionization parameter (e.g., \citealt{Cameron2023a}; \citealt{Sanders2023}), a low metallicity deviating from the mass-metallicity-SFR relationship (e.g., \citealt{Curti2023b}; \citealt{Nakajima2023}; \citealt{Tacchella2023}), and peculiar N/O and C/O (e.g., \citealt{Isobe2023}).
% At the low-metallicity regime, of primary nitrogen production, the N/O ratio is independent of metallicity
Therefore, the applicability of the nebular models needs to be extended to galaxies of a wider redshift range. 
    
There are various types of ionizing sources, including young massive stars, AGNs, post-asymptotic giant branch (post-AGB) stars, X-ray binaries, shocks, and (possibly) Pop III stars, and most realistically, a mixture of the above. 
%Among these, star-forming galaxies are proposed to be the dominant ionizing source in the early universe (e.g., \citealt{Hirschmann2023b}). 
A classical approach to diagnosing the nature of the ionizing sources is through UV and optical emission line ratios (e.g., \citealt{Kewley2019}; \citealt{Plat2019}; \citealt{Hirschmann2023b}). But such emission line ratio diagrams are often only useful at differentiating a few ionizing sources, such as star-forming galaxies and AGNs, where the other sources may not show a clear pattern. An additional challenge is that since the physical conditions driving the line ratios can evolve with redshift, the diagnostic criteria have to be adjusted for objects at different redshifts (e.g., \citealt{Steidel2014}; \citealt{Strom2017}; \citealt{Garg2022}). 

Additionally, the ionizing properties of these sources are often themselves uncertain. For example, models of massive stars are highly uncertain %Stellar models have significant uncertainties in their ionizing spectra particularly at the massive end, 
since observational constraints are limited due to their short lifetimes. The ionizing emission from massive stars is difficult to model as their evolutionary path and properties depend on the details of mass loss (e.g., \citealt{Smith2014}; \citealt{Steidel2016}; \citealt{Senchyna2021}), and factors like rotation and multiplicity can affect how long they live, how many ionizing photons they emit, and also the distribution of their ionizing photons as a function of wavelength (e.g., \citealt{Choi2017}). These model assumptions are hard to test directly from the observed photometry or spectrum because galaxy properties including age, stellar metallicity, SFR, and assumed initial mass function (IMF) will all influence the stellar ionizing spectrum. Another way to constrain the stellar models is through the nebular emission. The relative strengths of the emission lines provide information on the distribution of energy deposited into the gas around stars. While it is challenging to find a single line ratio as a definitive feature to differentiate stellar models, we can take advantage of multiple line fluxes to infer the full ionizing spectrum shape. Such nebular emission spectrum fitting requires a flexible nebular model for describing the emission line fluxes from different stellar models. 
%To describe the full nebular emission we can emulate the photoionization model to infer the full ionizing spectrum shape and constrain the stellar models directly.

Tremendous efforts have been invested in developing nebular emission models to interpret the emission line observations (e.g. \citealt{Steidel2016}; \citealt{Gutkin2016}; \citealt{Byler2017}). Many nebular emission models are based on photoionization modeling codes, e.g. \cloudy\ (\citealt{Ferland1998}; \citealt{Cloudy2023}) or MAPPINGS \citep{Sultherland2018} (e.g., \citealt{Charlot2001}; \citealt{Groves2004}; \citealt{Gutkin2016}; \citealt{Feltre2016}; \citealt{Morisset2016}; \citealt{Steidel2016}; \citealt{Byler2017}; \citealt{Berg2021}; \citealt{Umeda2022}). These works usually involve building static grids of the nebular model, by running the photoionization codes with different parameters many times and then interpolating between the grid values to generate appropriate models. Such nebular models have been further integrated into spectral energy distribution (SED) fitting frameworks to estimate the stellar and nebular properties self-consistently, such as \prospector{} \citep{prospector}, \texttt{BAGPIPES} \citep{Carnall2018}, \texttt{BEAGLE} \citep{Chevallard2016}, etc. 
Grid-based nebular models are accurate in learning the detailed physics of individual objects. However, because the number of nebular continuum and lines stored in the disk grows exponentially with the dimension of the grid, the allowed model space is limited by the memory. Also, interpolation within a large grid is memory-intensive. Hence, such models are usually built specifically to model certain types of ionizing sources such as star-forming galaxies or AGNs. To develop a general tool for interpreting the nebular emission powered by different sources, a higher-dimensional nebular model is necessary. 
%In this work, we employ neural networks to develop a 12-dimension nebular model.
%Much progress has also been made in building spectral energy distribution (SED) fitting frameworks to estimate the stellar and nebular properties self-consistently, such as \prospector{} \citep{prospector}, \texttt{BEAGLE} \citep{Chevallard2016}, etc. Integrating nebular emission modeling into SED fitting codes allows fitting the nebular and the stellar population properties in a more consistent way than inferring nebular conditions directly from line ratios. 

Due to both an increasingly complex and well-measured set of observations and to both speed-ups and grid compression in advanced machine learning architectures, now is a good time to build more flexible nebular emission models. In this paper, we present a neural net emulator around \cloudy\ with freedom in the ionizing spectrum shape and the ionizing photon input, gas density, gas-phase metallicity, [N/O] and [C/O] ratio. The wide coverage in the nebular parameter space makes this tool suitable for modeling a variety of ionizing sources and nebular conditions across different redshifts, and it allows the user to infer an ionizing spectrum instead of relying on fixed ionizing spectrum, e.g. a stellar ionizing spectrum determined by pre-computed stellar models. %a set of stellar isochrones and spectral library. 
Our neural net emulator also offers speed advantages, facilitating broad applications to large surveys. The paper is structured as follows.  Section~\ref{sec:cloudy} is about our \cloudy\ setups for modeling the HII regions, including the free parameters for describing the ionizing gas properties. In Section~\ref{sec:powerlaw}, we introduce our approximation of the ionizing spectrum and test and justify our approximation. In Section~\ref{sec:nn}, we describe the architecture and training process for the neural net emulator. We then conduct recovery tests in Section~\ref{sec:test} to evaluate the emulator's performance with mock emission line observations of different signal-to-noise ratios. Finally, in Section~\ref{sec:discussion}, we discuss the potential applications and limitations of our tool, providing an example of using \cue\ to distinguish mock ionizing spectra from different stellar models.

\section{\cloudy\ settings}\label{sec:cloudy}
We employ the spectral synthesis code \cloudy\ (version 22.00; \citealt{Cloudy2023}) to calculate the continuum and line emission from a single HII region ionized by a point source at the center. We largely adopt the \cloudy\ settings from \citet{Byler2017}, though we introduce greater flexibility in prescribing the ionizing radiation and the physical properties of the gas cloud. \cloudy\ takes as input the ionizing radiation striking the cloud, gas density, and the chemical composition and dust content of the gas, and computes the nebular continuum and line emission. For the line prediction, we adopt a line list of 128 emission lines from UV to far-infrared provided by \citet{Byler2017}. All line radiative transfer processes are included, such as recombination, collisional excitation, and collisional ionization (e.g., \citealt{Ferland2017}; \citealt{Tacchella2022}).
%line trapping, collisional deexcitation, continuum pumping, and destruction by background opacities.

Our model assumes a spherical shell gas cloud geometry. \cloudy\ solves for the ionization, density, and temperature structure across the spherical layers. The distance from the central source to the inner face of the cloud is fixed at $R_\mathrm{inner} = 10^{19}$\,cm following \citet{Byler2017}. We assume a covering factor of 1. We do not consider any escape of the ionizing radiation to the circumgalactic medium, which degenerates with the normalization of the ionizing spectrum and is effectively considered in our model as we let normalization vary. %which could be important for certain lines at high redshift.

For the chemical composition of the ionizing gas, we allow freedom in the gas-phase metallicity specified by [O/H], [C/O], and [N/O]. The gas-phase metallicity usually correlates with the stellar metallicity in galaxies (e.g., \citealt{Gallazzi2005}) but effects such as pristine gas inflow can dilute the metallicity of the gas where the stars live in. We scale element abundances linearly with (O/H)/(O/H)$_\odot$, the oxygen abundance relative to solar by number, with the exception of helium, carbon, and nitrogen. The He abundance is drawn from a linear relationship with metallicity following \citet{Dopita2000}. Nitrogen and carbon production have secondary production mechanisms and their relationships with [O/H] are complicated and long debated  (e.g., \citealt{Akerman2004}; \citealt{Groves2004}; \citealt{Shapley2015}; \citealt{Nicholls2017}; \citealt{Berg2019}; \citealt{Isobe2023}).
Hence, we treat [C/O] and [N/O] as free parameters in our nebular emulator.  Additionally, we apply constant dust depletion factors $D$ of metals. The solar abundances and dust depletion are specified by \citet{Dopita2000}. For reference, the solar values we adopted are $\log(\mathrm{O}/\mathrm{H})_\odot = -3.07$, $\log(\mathrm{C}/\mathrm{O})_\odot = -0.37$, $\log(\mathrm{N}/\mathrm{O})_\odot = -0.88$, and $\log D_\mathrm{O} = -0.22$, $\log D_\mathrm{C} = -0.30$, $\log D_\mathrm{N} = -0.22$. 

To derive the ionization structure, a crucial input is the ionization parameter $U \equiv \left(n_\gamma / n_{\mathrm{H}}\right)$. Here $n_\gamma$ is the number density of the ionizing photons, and $n_{\mathrm{H}}$ is the number density of the hydrogen. $U$ characterizes the strength of the ionization field and is defined as the ratio of the isotropic ionizing radiation from a central source to the gas density
\begin{equation}
    U = \frac{Q_\mathrm{H}}{4 \pi R_\mathrm{inner}^2 n_\mathrm{H} c}. 
\end{equation}
In our emulator, $U$ is unfolded into two free parameters, the hydrogen ionizing photon rate $\QH = \frac{1}{hc} \int_{0}^{912\mathrm{\AA}} \lambda F_\lambda \mathrm{d}\lambda$, and $\nH$. We assume a constant $\nH$ across the HII region, and this assumption will be discussed in Section~\ref{sec:discussion}. When analyzing the nebular emission from observed galaxies, there are certain cases we want to link the stellar populations produced by the SED fitting codes to the nebular emission and force the ionizing spectrum to be the stellar spectrum. In a such situation, we will calculate the effective $\QH$ from the stellar continuum and scale the nebular emission according to the ratio of the effective $\QH$ and the inferred $\QH$ from the emulator. In this way, we effectively view the ionizing region of galaxies as the sum of multiple HII regions of the same gas properties.  

The final component of the nebular model, the incident ionizing radiation, is approximated as a piece-wise continuous 4-part power-law. In the next section, we will describe in detail how we choose the wavelength segments of the power-law approximation and present the robustness tests for this approximation. Since $\QH$ specifies the normalization of the ionizing spectrum, we can reduce the free parameters for describing the 4-part power-laws from eight to seven. That being said, our free parameters for characterizing the ionizing spectrum shape are the four power-law indexes $\alpha$, and the ratios between the integrated fluxes $F$ of each piecewise power-law. 

%Discuss the relationship between nH, gas density and electron density
%For the rest elements, their abundances are log (abundance) = log Z + log(abundances from Dopital2001) + depletion factor

In summary, we vary the ionizing spectrum, $\QH$, $\nH$, [O/H], [C/O], [N/O] of the HII region, leading to 12 free parameters in total, and run \cloudy\ to compute the emitted continuum and line emission. The 12 free parameters and their range are listed in Table~\ref{tab:cloudy_prior}. We set the range of the parameters describing the ionizing spectrum such that they cover all types of sources in Figure~\ref{fig:ion_spec} (see the description for these ionizing sources in Section~\ref{sec:powerlaw}). The upper and lower limits of the parameters describing nebular properties are adopted from \citet{Gutkin2016} and \citet{Byler2017}.

Despite this new flexibility, we must still make specific assumptions and fix certain parameters such as $R_\mathrm{inner}$ in the nebular model for practical purposes. Introducing additional degrees of freedom may increase the size of the neural network emulator, potentially compromising computational speed, or it may necessitate limiting the complexity of the neural network, which could impact accuracy. In addition, some fixed nebular parameters are degenerate with our free parameters and could be challenging to model.

\begin{deluxetable*}{ccc}
    \tablenum{1}
    \tablecaption{Free parameters describing the ionizing spectrum and the nebular properties.}\label{tab:cloudy_prior}
    \tablewidth{0pt}
    \tablehead{
    \colhead{Parameter} & \colhead{Description} & \colhead{Range} %\multicolumn2c{}
    }
    %\decimalcolnumbers
    \startdata
    $\alpha_\mathrm{HeII}$ & power-law slope at $1\,\mathrm{\AA}<\lambda<228\,\mathrm{\AA}$ & [1, 42]\\
    $\alpha_\mathrm{OII}$ & power-law slope at $228\,\mathrm{\AA}<\lambda<353\,\mathrm{\AA}$ & [-0.3, 30]\\
    $\alpha_\mathrm{HeI}$ & power-law slope at $353\,\mathrm{\AA}<\lambda<504\,\mathrm{\AA}$ & [-1.1, 14]\\
    $\alpha_\mathrm{HI}$ & power-law slope at $504\,\mathrm{\AA}<\lambda<912\,\mathrm{\AA}$ & [-1.7, 8]\\
    $\log {F_\mathrm{OII}}/{F_\mathrm{HeII}}$ & flux ratios of the two bluest segments & [-0.1, 10.1]\\
    $\log {F_\mathrm{HeI}}/{F_\mathrm{OII}}$ & flux ratios of the second and the third segments & [-0.5, 1.9]\\
    $\log {F_\mathrm{HI}}/{F_\mathrm{HeI}}$ & flux ratios of the two reddest segments &[-0.4, 2.2]\\
    \hline
    $\log U$ & ionization parameter & [-4, -1]\\
    $\log \nH (\mathrm{cm}^{-3})$ & hydrogen density& [1, 4]\\
    $\log (\mathrm{O}/\mathrm{H})/(\mathrm{O}/\mathrm{H})_\odot$ & oxygen abundance & [-2.2, 0.5]\\
    $(\mathrm{C}/\mathrm{O})/(\mathrm{C}/\mathrm{O})_\odot$ & carbon-to-oxygen ratio & [0.1, 5.4]\\
    $(\mathrm{N}/\mathrm{O})/(\mathrm{N}/\mathrm{O})_\odot$ & nitrogen-to-oxygen ratio & [0.1, 5.4]\\
    %\hline
    \enddata
    \tablecomments{The ionizing spectrum is segmented into a 4-part power-law (see Equation (\ref{eq:powerlaws})). The top seven parameters control the shape of the ionizing spectrum. The bottom five parameters characterize the ionizing gas properties. The third column specifies the allowed range of the parameters in the training set.}
\end{deluxetable*}

\section{power-law approximation of the ionizing spectrum}\label{sec:powerlaw}
%JWST has revolutionized the field of extragalactic astronomy with its sensitive and high-resolution infrared view of the distant universe.
A flexible model of the ionizing spectrum is crucial for interpreting the complex nebular emission properties of galaxies at different redshifts and in different environments, such as the extreme emission line properties of the high-redshift sources (e.g., \citealt{Williams2023}; \citealt{Bunker2023}). In this section, we will present our approach to incorporating the ionizing spectrum of various types of sources into our nebular model. To achieve this goal, we approximate the ionizing spectrum with a piecewise power-law. In this way, we are agnostic to the actual physics of the ionizing source, but we tune the allowed range of this piecewise power-law to allow a few specific types of sources.

\subsection{The ionizing spectra of different astrophysical sources.}
\begin{figure*}
    \centering
    \includegraphics{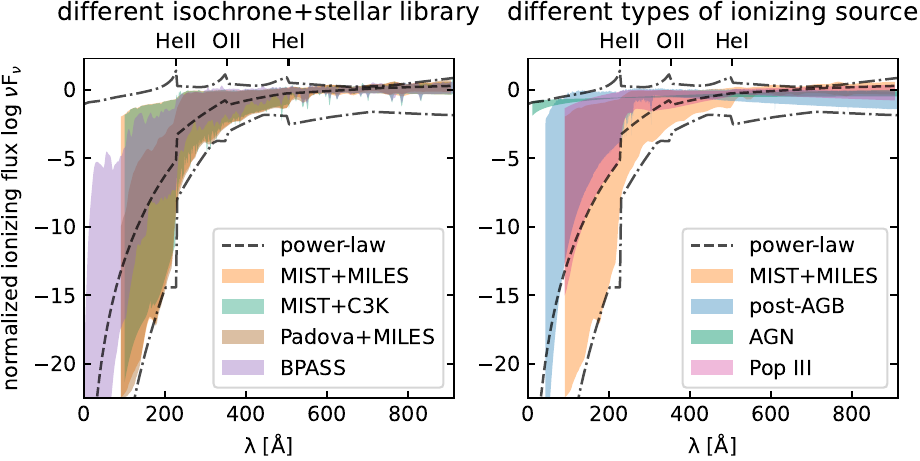}
    \caption{Here we illustrate the range of ionizing spectra from different sources that can be modeled using the training set. The left panel compares the range of ionizing spectra over many stellar ages and metallicities, in different stellar isochrones and spectral libraries. The right panel compares the range of ionizing spectra from SSPs, post-AGB stars, AGNs, and popIII stars. The black dashed line represents the median of our 4-segment power-law ionizing spectrum in the training set, and the lower and upper limits of the training ionizing spectra are in dashed-dotted lines. The spectra are normalized by the total ionizing luminosity. The ionizing spectrum shape provides insights into the type of ionizing sources and the stellar isochrones and libraries used in SED modeling. Our nebular emulator describes the input ionizing spectrum with a flexible 4-part continuous power-law to encompass various types of ionizing spectra.}
    \label{fig:ion_spec}
\end{figure*}

We design the allowed range of our ionizing spectrum parameterization to incorporate various types of astrophysical sources shown in Figure~\ref{fig:ion_spec}. In the left panel of Figure~\ref{fig:ion_spec}, we depict the ionizing spectra of simple stellar populations (SSPs) based on different stellar models and libraries (MIST+MILES, MIST+C3K, BPASS, Padova+MILES), all assuming a \citet{Kroupa2001} IMF. The stellar models and isochrones are compiled through the Flexible Stellar Population Synthesis (FSPS; \citealt{Conroy2009, Conroy&Gunn2010}) framework. Only SSPs younger than 25\,Myr are shown here due to their dominant contribution to the total ionizing budget compared to old SSPs. Older SSPs are also more difficult to parameterize due to their complex spectrum shapes, and as a result, may not be well fitted by the emulator. However, in practice, for most applications this will be a minor concern since the young SSPs dominate the contribution to the total ionizing radiation from star-forming galaxies, particularly at high redshifts.

Notably, different stellar models yield substantially different ionizing spectra as illustrated in Figure~\ref{fig:ion_spec}. For example, single-star evolution model Padova \citep{Girardi2002} and binary stellar evolution model BPASS (v2.2; \citealt{bpass}) produce different amounts of ionizing photons and the shape of their ionizing spectra are significantly different in the blue end. MIST isochrones \citep{Choi2016} take into account the effect of rotation, which can prolong ionizing photon production and also produce a harder ionizing spectrum (e.g., \citealt{Leitherer2014}; \citealt{Choi2017}). These variations underscore the potential of using nebular emission to directly interrogate different stellar models. We will further demonstrate in Section~\ref{sec:discussion} that, when given UV--optical spectra with reasonable observational uncertainties,  our nebular model is able to differentiate single and binary stellar evolution models.

Distinct sources also produce unique ionizing radiation.
%Nebular emission line ratios are common diagnostics of ionizing sources. For example, star-forming galaxies and AGNs dominate different regions in the BPT diagram. The underlying reason for the distinct emission line ratios among sources is their different ionizing spectrum. 
In the right column of Figure~\ref{fig:ion_spec}, we present the range of ionizing spectra of SSPs, post-AGB stars, AGNs, and PopIII stars. Post-AGB spectra are from \citet{Rauch2003}. AGN ionizing spectra are assumed to follow a power-law $F_\nu = A \lambda^{\alpha}$, with the power-law index $1.2 \le \alpha \le 2$ (e.g., \citealt{Groves2004}; \citealt{Feltre2016}). Pop III star spectra are from \citet{Schaerer2002}\footnote{The ionizing spectra for PopIII stars are clipped at a minimum of $10^{-70}$ erg/s/Hz/$\mathrm{L}_\odot$.}, which are purely theoretical and highly uncertain, but included here for reference. This figure highlights the distinct ionizing spectral shapes of these sources, which in turn allows us to use observed nebular emission lines to diagnose the nature of ionizing sources. For example, AGNs have a hard ionizing spectrum in the blue end, leading to strong high ionization state lines, whereas SSPs have redder ionizing spectra and present weaker high ionization lines.

\subsection{Segmenting and fitting ionizing spectra with a piecewise power-law}
We seek to build a versatile and general nebular model capable of describing the ionizing spectrum ($1\,\mathrm{\AA} \leq \lambda \leq 912\,\mathrm{\AA}$) of various sources, including less understood sources at high redshift, old ionizing spectra, etc. 
Examination of ionizing spectra of stellar SSPs, post-AGBs, PopIII stars, and AGNs in Figure~\ref{fig:ion_spec} indicates that they loosely follow power-laws with sharp ionizing edges.

We segment the ionizing spectrum based on ionization edges and fit a power-law to the spectrum in each part. The selection of segment boundaries is critical for 
%for delineating the ionizing spectrum shape and 
accurately reproducing the nebular emission, and thus, choosing ionization edges right is critical to getting the temperature structure right. The temperature structure of the nebula depends on the energy deposited across the HII region, making emission lines, especially fine structure lines sensitive to the ionization continuum shape. Most stellar ionizing spectra exhibit a prominent discontinuity at the HeII ionization edge. The HeI ionization edge also appears to be a discontinuity in many SSPs and is important for tracking the He ionization structure. Furthermore, our experiments indicate that introducing a cut at the OII ionization edge significantly enhances the goodness of fit for the far-infrared fine structure lines. In summary, we choose the boundaries of each segment to be the ionizing potential of HeII (228\AA), OII (353\AA), and HeI (504\AA) based on our experimentation. 
%The ionizing spectra have been normalized by its median to avoid numerical issues. 
%The ionization structure can be very sensitive to $T_e$ when $T_e$ is in the transition region.

We employ a 4-part power-law fit to the ionizing spectrum, where the flux $F_{\nu}$ is given by 
\begin{equation}\label{eq:powerlaws}
    F_\nu = 
    \begin{cases}
    A_\mathrm{HeII} \lambda^{\alpha_\mathrm{HeII}} & \mathrm{for}~1\,\mathrm{\AA}<\lambda<228\,\mathrm{\AA};
    \\ A_\mathrm{OII} \lambda^{\alpha_\mathrm{OII}} & \mathrm{for}~228\,\mathrm{\AA}<\lambda<353\,\mathrm{\AA};
    \\ A_\mathrm{HeI} \lambda^{\alpha_\mathrm{HeI}} & \mathrm{for}~353\,\mathrm{\AA}<\lambda<504\,\mathrm{\AA};
    \\ 
    A_\mathrm{HI} \lambda^{\alpha_\mathrm{HI}} & \mathrm{for}~504\,\mathrm{\AA}<\lambda<912\,\mathrm{\AA}.
    \end{cases}
\end{equation}
As introduced in Section~\ref{sec:cloudy}, we use the total flux ratios between the adjacent segments ${F_\mathrm{OII}}/{F_\mathrm{HeII}}$, ${F_\mathrm{HeI}}/{F_\mathrm{OII}}$, ${F_\mathrm{HI}}/{F_\mathrm{HeI}}$ instead of the four normalization parameters $A_\mathrm{HeII}$, $A_\mathrm{OII}$, $A_\mathrm{HeI}$, and $A_\mathrm{HI}$ to describe the normalization of each segments. Combining with $\QH$ and $\alpha$, the flux ratios can be converted back to the normalization parameters.

Our fit to the ionizing spectrum flux needs to match both the spectrum shape and the total photon production rate $Q = \int_{\nu_\mathrm{min}}^{\nu_\mathrm{max}} \frac{F_\nu}{h\nu} \mathrm{d}\nu$ of each segment. A typical linear regression assigns equal weights to each data point, which does not preserve $Q$. Therefore, we customize our loss function for power-law fits to include the fit error of both $F_{\nu}$ and $Q$. The loss function of the power-law fit to one segment becomes 
\begin{flalign}\label{eq:loss}
    \mathcal{L} = &\sum_N{(\log F_{\nu, \rm{true}} - \log F_{\nu, \rm{pred}})^2} + \\\nonumber
    &N (\log Q_{\rm{true}} - \log Q_{\rm{pred}})^2. \nonumber
\end{flalign}
Here $F_{\nu, \rm{true}}$, $Q_{\rm{true}}$ are the spectral flux and total photon production rate for our training ionizing spectra respectively. $N$ is the number of spectral flux points in each bin. The photon rates at the four segments are dubbed $Q_\mathrm{HeII}$, $Q_\mathrm{OII}$, $Q_\mathrm{HeI}$, $Q_\mathrm{HI}$ for convenience. As a reminder, these are not the traditional definitions of $Q$, unlike our definition of $\QH$ in Section~\ref{sec:cloudy}.

We fit the ionizing spectra of different sources shown in Figure~\ref{fig:ion_spec} to determine the allowed range of the input power-law parameters to our nebular emission emulator (see Section~\ref{sec:cloudy}). Restricting the range of parameters is important to produce a higher neural network accuracy. We will create a sample of random spectra in this range for training purposes in Section~\ref{sec:nn}. As shown in Figure~\ref{fig:ion_spec}, some spectral libraries do not have wavelength grid points below 100\,\AA. For these SSPs, we simply fit the power-laws to the available spectrum grids and extrapolate the fit to 1\,\AA. It is a reasonable choice since we do not model very high ionization state lines with ionization edge below 100\,\AA. In the context of stellar populations, ionizing radiation at $\lambda < 100$\,\AA\ has a marginal impact on for the emulated emission lines. For example, it contributes $\lesssim 0.1$\% $Q_\mathrm{HeII}$ for BPASS SSPs. % and highly uncertain

Because UV and optical light can ionize certain metal ions, our decision to model only the hydrogen ionizing part will not perfectly describe the ionization structure for these metals and may affect the goodness of fit for the continuum and line emission. Therefore, we place a threshold $R_\mathrm{ionizing} = \log\left(F_{\lambda<912}/F_{912-7000}\right)$ on the ionizing spectra included in the training set with $R_\mathrm{ionizing} > -1.5$. %Q$_\rm{H}$ cut at $10^{45}$ s$^{-1}$ per $\Msun$ formed. 
This cut is a reasonable assumption in reproducing the overall nebular emission in galaxies because the ionizing spectra with small $R_\mathrm{ionizing}$ usually generate fewer ionizing photons and contribute less to powering the total nebular emission from the galaxy. We also extrapolate the last piece of power-law at 504--912\,\AA\ redwards to 2000\,\AA\ to account for some level of UV radiation when generating \cloudy\ inputs for the training set. We will discuss the effect of this extrapolation in Section~\ref{subsec:powerlaw_test}.

\subsection{Accuracy of the power-law approximation}\label{subsec:powerlaw_test}

\begin{figure*}
    \centering
    \includegraphics[width=0.9\textwidth]{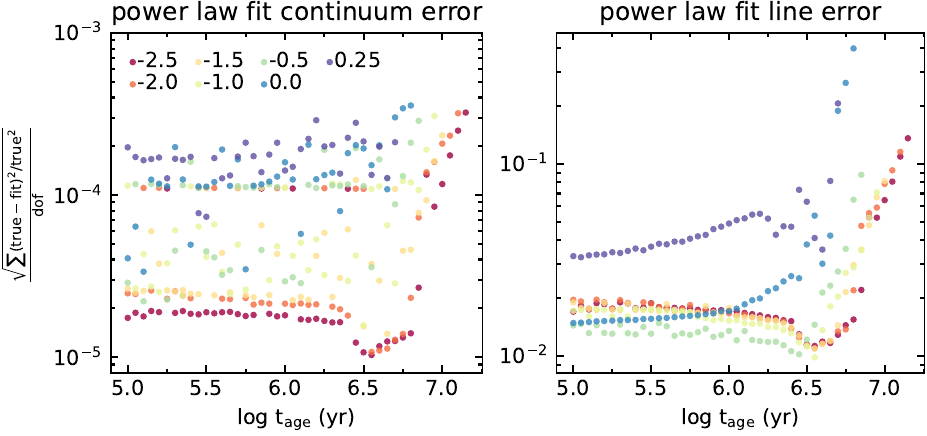}
    \caption{Nebular continuum and line emission errors due to approximating the ionizing SSPs with a 4-segment power-law as a function of the SSP ages, color-coded by the SSP metallicities. The SSPs are generated with the MIST isochrones and MILES library. Only SSPs producing strong ionizing flux ($R_\mathrm{ionizing} > -1.5$) are included. The gas-phase metallicity is assumed to be the same as the stellar metallicity. Left: $\chi^2$ fractional difference between the nebular continuum powered by SSPs and the ones powered by the power-laws from the CLOUDY runs. Right: $\chi^2$ fractional error of the emission line luminosities, representing the average fractional errors of all 128 emission lines in our line list. The power-law approximation to the ionizing SSPs introduces $<$1\textperthousand{} errors in the continuum emission and $\lesssim$10\% errors in the line emission calculated by CLOUDY. The approximation is less accurate for older and higher metallicity SSPs.}
    \label{fig:powerlaw_ssp_error}
\end{figure*}

\begin{figure*}
    \centering
    \includegraphics[width=0.9\textwidth]{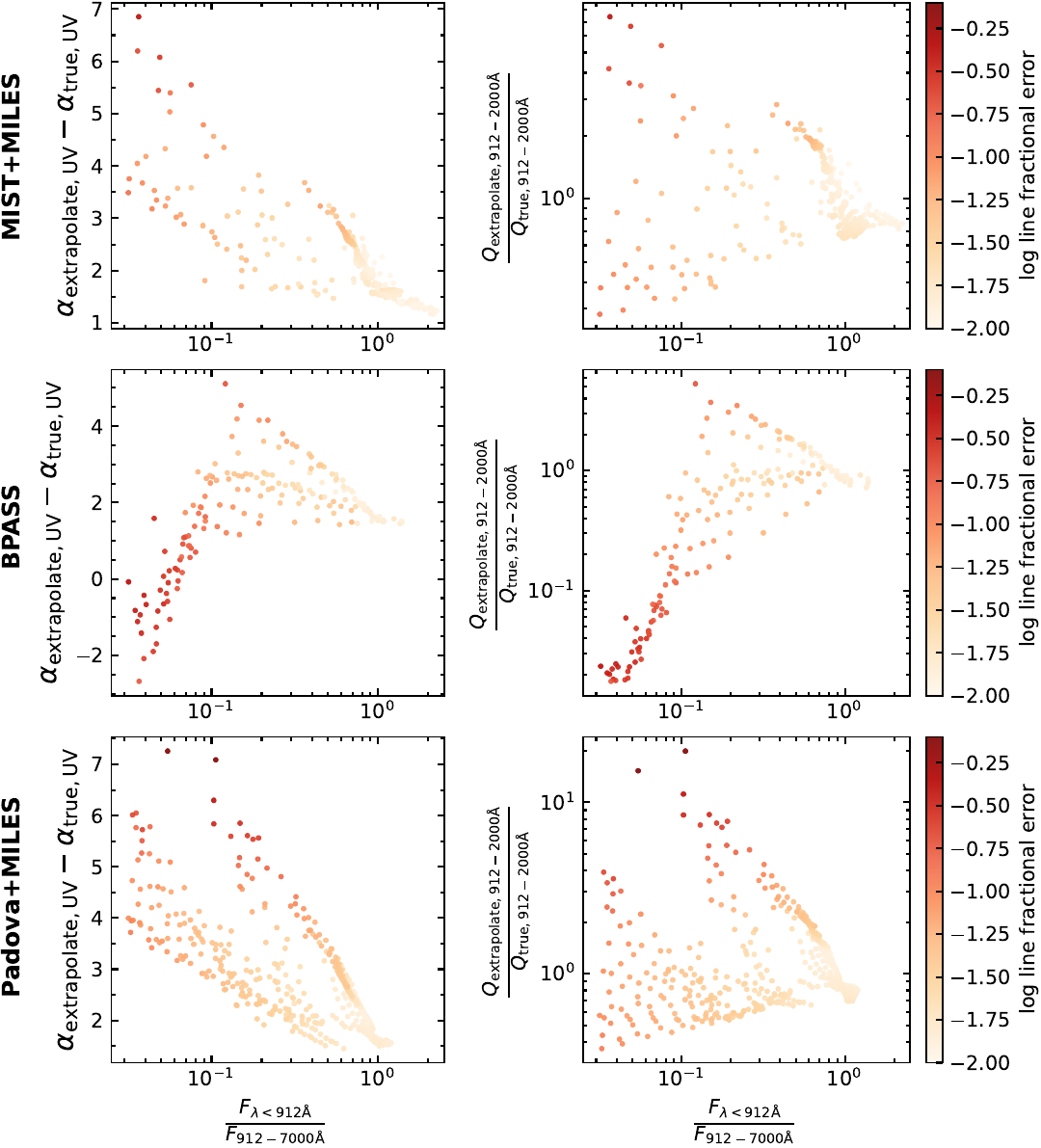}
    \caption{The emission line error of our power-law approximation is largely driven by the discrepancy between the actual UV spectrum of SSPs and our extrapolation of the 504--912\,\AA\ power-law fit to the UV. In the left column, we illustrate the relationship between the average $\chi^2$ fractional error of the emission lines and ionizing-to-UV-optical fraction $R_\mathrm{ionizing}$ of an SSP, along with the UV index error due to the extrapolation.  The right column demonstrates the emission line errors as a function of $R_\mathrm{ionizing}$ and the ratio of the UV photon rate from the extrapolation and the true UV photon rate. Only SSPs with $R_\mathrm{ionizing} > -1.5$ are shown here. Our parametrization of the ionizing spectrum introduces larger fractional errors on the predicted emission line luminosities when the UV photons from the power-law extrapolation are off by a factor of $\gtrsim 3$. Additionally, the average fractional emission line error increases as $R_\mathrm{ionizing}$ decreases.}
    \label{fig:UV_fit}
\end{figure*}

We assess the robustness of the power-law approximation to the ionizing spectrum by comparing their \cloudy\ outputs. To be more specific, we run actual stellar ionizing spectra through \cloudy, and also power-law approximations of these spectra, then compare the resulting nebular continuum and line emission. Our tests encompass ionizing spectra of SSPs and composite stellar populations (CSPs). The SSP tests provide a more direct reflection of the trends with the stellar population properties, while CSP tests align more closely with the real situation in galaxies, where HII regions can be ionized by stars with a wide range of ages.

The results for SSPs are presented in Figure~\ref{fig:powerlaw_ssp_error}. We illustrate the average fractional error of continuum and line emission introduced by the power-law parameterization at each SSP age and metallicity. The SSPs are generated with the MIST isochrones and MILES library. Only SSPs with $R_\mathrm{ionizing} > -1.5$ are included. In this test, we assume the same gas-phase metallicity as the stellar metallicity, solar abundance ratios, ionization parameter $\log U = -2.5$, and gas density $\nH = 100$\,$\mathrm{cm}^{-3}$. 

The continuum error introduced by power-law approximation is less than 1\textperthousand{}, with larger errors for older and higher metallicity SSPs. Older SSPs have more absorption and emission features in their ionizing spectrum, with smaller $R_\mathrm{ionizing}$ than young SSPs, producing more UV photons that are not captured by the power-laws but still affect the nebular structure. The power-law approximation is also less perfect for the metal-rich SSPs due to their more complex ionizing spectra shape.

\begin{figure}
    \centering
    \includegraphics{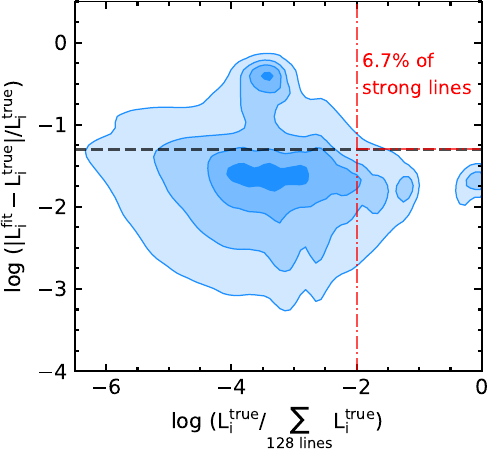}
    \caption{Errors of individual lines raised by the power-law approximation to CSPs against the contribution of each line to the total line luminosities. Red dashed lines demarcate the strong lines with large errors. The power-law approximation introduces $< 5$\% errors for most strong lines.   %(maybe colored by metallicity and ionization states)
    }
    \label{fig:line_error}
\end{figure}

Similar arguments apply to the line estimates. The emission lines from the power-law fits are consistent with those from SSPs within 10\%, with increasing errors for older SSPs and the largest errors for $\log Z/Z_\odot = 0.25$ SSPs. The average emission line errors shown here are dominated by weak lines as they are more sensitive to the detailed shape of the ionizing spectrum. In practice, the strong lines are easier to observe and more important for our emulator predictions. The power-law fit errors of individual lines will be explored in Figure~\ref{fig:line_error}.

We further evaluate the accuracy of the power-law approximation to SSPs generated from different stellar models in Figure~\ref{fig:UV_fit}. In particular, we examine the accuracy of extrapolating the reddest power-law into the UV. Compared to the SSPs generated with the MIST isochrones and MILES library, the results for BPASS and Padova stellar populations show larger emission line offsets. Given that the power-law approximations to SSP ionizing spectra from different stellar models show consistent goodness of fit, one possibility is that the UV radiation drives the larger emission line errors for BPASS and Padova instead of the fit quality of the ionizing spectrum. To explore this hypothesis, we compare the UV photons predicted by the extrapolated power-law to the actual UV emission from the SSPs and show their ratios in the right column of Figure~\ref{fig:UV_fit}. In our extrapolation, the UV slope $\alpha_\mathrm{extrapolated, UV}$ by definition is $\alpha_\mathrm{HI}$. We fit an additional power-law to the SSPs at 912--2000\,\AA\ and compare this UV slope $\alpha_\mathrm{true, UV}$ to $\alpha_\mathrm{HI}$ in the left column of Figure~\ref{fig:UV_fit}.

According to Figure~\ref{fig:UV_fit}, the power-law approximation can get notably worse depending on the discrepancy between the extrapolated reddest ionizing spectrum and the actual SSPs. This discrepancy has a greater impact when the UV-optical spectrum of an SSP is important relative to the ionizing spectrum, as indicated by a small $R_\mathrm{ionizing}$, e.g. for older stellar populations. Our approximation works well for most MIST and Padova SSPs, while a large portion of BPASS SSPs exhibit average emission line error $\gtrsim 10$\%. This is specifically because the extrapolation for older BPASS SSPs tends to under-predict the UV photons with a flatter UV slope than the true one, particularly when $R_\mathrm{ionizing} < 0.1$. Conversely, the extrapolation for young and metal-rich BPASS SSPs tends to over-predict the UV photons with a steeper UV slope than the true slope. On the other hand, the power-law extrapolation consistently yields a redder UV slope than the actual MIST SSPs and Padova SSPs. Consequently, the emission lines exhibit the greatest offsets when the extrapolation predicts the highest number of UV photons. We find that this extrapolation error of the UV spectrum primarily affects weak emission lines and those at wavelengths below $2000$\,\AA, notably hydrogen lines blueward of Ly$\alpha$, [CIII] 1909\,\AA, [O III] 1661\,\AA, and [O III] 1666\,\AA.
%[Al III] 1863\,\AA, [Al III] 1855\,\AA[O III 1661\,AA, [O III] 1666\,AA
In future work, we may extend our power-law parametrization to the UV to achieve a more accurate emulator.

\begin{figure*}
    \centering
    \includegraphics[width=0.8\textwidth]{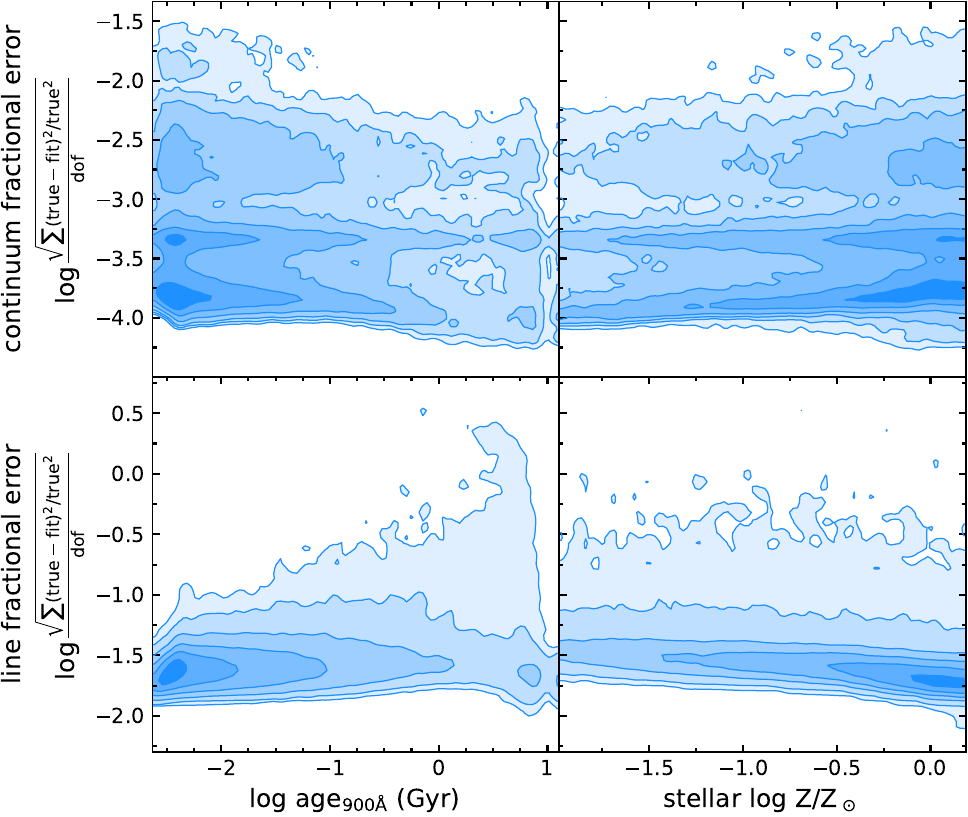}
    \caption{Errors on the nebular emission raised by the 4-segment power-law approximation to the ionizing CSPs as a function of the light-weighted age at 900\AA\ and stellar metallicity. The CSPs are generated with MIST isochrones and C3K library. Top panel: $\chi^2$ fractional errors of the nebular continuum. Bottom panel: $\chi^2$ fractional error of the emission line luminosities. The power-law approximation to the ionizing CSPs introduces $<$1\textperthousand{} errors in the continuum emission and $\lesssim$10\% errors in the line emission calculated by CLOUDY. The power-law approximation is marginally less accurate for older and lower metallicity CSPs.}
    \label{fig:powerlaw_csp_error}
\end{figure*}

In Figure~\ref{fig:powerlaw_csp_error}, we depict the power-law approximation errors for CSPs based on the MIST isochrones and C3K library. These CSPs are generated at $z = 3$ from a flexible or ``nonparametric" method \citep{Leja2019a} which allows for a wide range of star formation histories (SFHs) using \prospector. We randomize the stellar metallicity, $U$, $\nH$, gas-phase metallicity [O/H], [C/O], and [N/O] to evaluate the robustness of the power-law approximation across the entire parameter space of our emulator. 

The continuum and line emission errors from fitting the CSPs show overall similar but weaker trends as the results for SSPs, with lower errors in the mean. There is a weak trend between the mean of the line errors and the CSP light-weighted ages, but the scatter of this relationship increases significantly compared to the SSPs. %e.g. all of the large line errors are at old ages
The average line errors slightly increase as the stellar metallicity decreases.
%, \textbf{potentially because the errors are dominated by weak lines and they are fainter at the metal-poor csps and more easily affected by the power-law approximation of the ionizing spectrum.} 
Apart from the CSP properties, the nebular parameters (e.g., nebular metallicity, etc.) also do not have significant trends with the power-law fit errors and thus, we do not show them here.

We further examine the accuracy of the power-law approximation for individual emission lines in Figure~\ref{fig:line_error}. This figure illustrates the strength of the emission lines compared to the line errors. It suggests that the power-law fit is valid for most strong emission lines. For the lines contributing more than 1\% of the total nebular emission line flux of each CSP, only 6.7\% have approximation errors $>5$\%. We find that the UV radiation ($\lambda > 912$\,\AA) not considered in the power-laws is the main reason for the large errors in weak lines, evidenced by the correlation between the individual line errors and the continuum errors. Since in practice, the strong lines will have the clearest observations, the large power-law approximation uncertainty of some weak lines will not have significant effects on the emulator efficacy when modeling real galaxy observations except for some specific cases, e.g. quiescent galaxies with some residual star formation. 

\section{A neural net emulation for CLOUDY}\label{sec:nn}

As we have laid out the ionizing radiation and nebular conditions \cue\ adopted as \cloudy\ inputs, in this section, we will introduce how we emulate \cloudy\ with neural nets and demonstrate the performance of the emulator. The emulator performance goal is to ensure the neural net uncertainties are smaller than current theoretical and observational uncertainties. We predict the nebular continuum and line emission powered by a set of SSPs using the 2013 and 2017 versions of \cloudy. Their offsets are around 5\%. Furthermore, spectroscopic slit losses from ground-based observations are typically $\gtrsim 1.5$ (see e.g., \citealt{Kriek2015}). That being said, our emulator does not need to be more precise than $\sim$5\%.

\subsection{Architecture and training of the neural net emulator}
With telescopes such as JWST, the Hubble Space Telescope (HST), Keck, and the Very Large Telescope (VLT) (for example, the upcoming instrument Multi-Object Optical and Near-infrared Spectrograph (MOONS)) accumulating a massive amount of high-quality emission line measurements, a fast, accurate, and universal framework for emulating the nebular emission is desirable and neural networks (NN) provide an excellent option. By training a large number of photoionization models into a compact neural network, we can achieve a significant speedup at evaluating the nebular emission over a broad model space. Also, its highly flexible and nonlinear structure allows the neural net to accurately describe the complex physics governing galaxy nebular emission.

We adopt the emulator architecture from \citet{Alsing2020}. The neural network consists of three hidden layers of 256 units each and a final output layer. Each hidden layer evaluates a nonlinear activation function on the fully-connected units (Equation 8; \citealt{Alsing2020}). The final layer uses a linear activation function. The training process involves four steps with learning rates of $10^{-3}$, $10^{-4}$, $10^{-5}$, and $10^{-6}$, and batch sizes of $10^{3}$, $10^{4}$, $5 \times 10^{4}$, and the size of the remaining training set respectively. The neural network is trained in \texttt{TENSORFLOW} with a mean squared error loss function and the stochastic gradient descent optimizer \texttt{ADAM}. 

We generate a training set of $2 \times 10^{6}$ sample, and a test set of $2 \times 10^{5}$ sample. 10\% of the full training dataset is devoted to validation. To generate the training and test set, we randomly choose the free parameters within the range listed in Table~\ref{tab:cloudy_prior} and use a pure piecewise power-law spectrum as input to \cloudy. We then run \cloudy\ to compute the nebular continuum and line emission based on the nebular model described in Section~\ref{sec:cloudy}. %The weights and biases

\subsubsection{PCA decomposition of the nebular continuum and line emission}
\begin{figure}
    \centering
    \includegraphics[width=0.95\columnwidth]{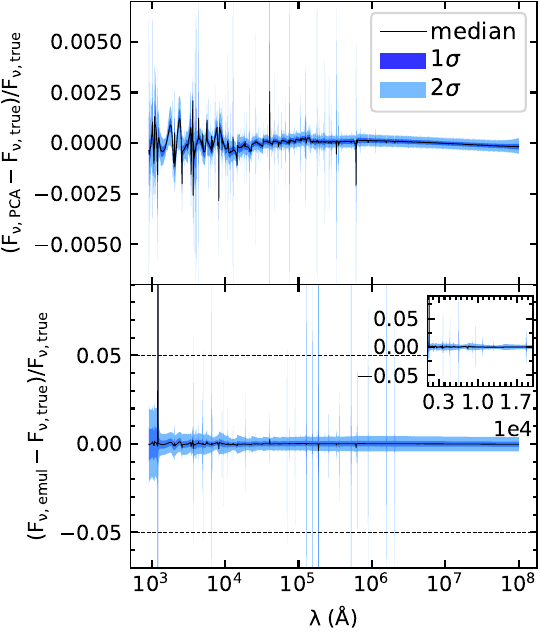}
    \caption{Nebular continuum emulator errors. The top panel shows the fractional errors of the PCA decomposition. The bottom panel shows the fractional errors of the continuum emulator, where the figure inside zooms into the emulator performance at optical--NIR wavelengths. The continuum emulator has a $\lesssim2\%$ uncertainty across all wavelengths.}
    \label{fig:cont_nn}
\end{figure}

Following the methodology of \citet{Alsing2020}, we perform a principal component analysis (PCA) decomposition on the nebular emission, using incremental PCA from the python package \texttt{SCIKIT-LEARN}. We subsequently train neural nets on the PCA basis coefficients, leading to a substantial reduction in the dimensionality of the training set. This allows us to have a small neural net architecture, and thus an acceleration of the emulator. %potentially better represents the physics

To construct the PCA basis of the nebular continuum, we first interpolate the nebular continuum from \cloudy\ outputs to the MILES wavelength grid to reduce the data size. We then perform the PCA decomposition on the interpolated nebular continuum. In Figure~\ref{fig:cont_nn}, we illustrate that 50 PCA components can represent the full 1000\,\AA--10\,mm continuum with an error of $<3$\textperthousand{}. This significant dimensionality reduction is expected since the nebular continuum at different wavelengths follows power laws, governed by atomic physics. 

The relationships between emission lines and nebular conditions are more intricate compared to the continuum, requiring a larger number of PCA basis components to adequately represent them. We group the emission lines according to elements and ionization potentials, as detailed in Section~\ref{subsubsec:line_emu}. PCA decomposition is then performed on each subgroup of lines, with the number of PCA components determined by the minimal requirements to achieve a PCA decomposition error of less than $<5$\%. In total, we utilize 76 PCA components to represent the 128 modeled emission lines.

\subsubsection{Continuum emulator}
The nebular continuum exhibits a generally smooth profile with distinct discontinuities. Unlike stellar continuum breaks, the nebular continuum demonstrates higher flux on the blue side of these discontinuities. This characteristic results in spectral features such as the Balmer jump, indicating strong nebular emission from the galaxy. 

The bottom panel of Figure~\ref{fig:cont_nn} shows the performance of the nebular continuum emulator. The plotted emulator errors originate from two sources: the PCA decomposition and the neural net approximation of the relationship between the input nebular physics and the output PCA coefficients. Across all wavelengths, we achieve an error of $\lesssim2\%$. Some larger emulator errors are noticeable around the discontinuities. These errors only have a small influence on the accuracy of the edge locations and do not affect the overall shape of the predicted nebular continuum.

\subsubsection{Line emulator}\label{subsubsec:line_emu}
We emulate a total of 128 emission lines using the line list provided by \citet{Byler2017}, with slight adjustments to match the vacuum wavelengths of the lines to the latest \cloudy\ version. Because the emission lines have diverse production mechanisms and originate from different ions, a single neural network to emulate all lines is impractical. Since the densities of ions depend on their chemical abundances, we first categorize lines according to the element species. Initially, we separate the emission lines into 5 groups of element species, H, He, C, N, O, and the others. Next, because ions of different ionization potentials respond differently to the change of electron temperature and probe unique regime of the ionizing spectrum (e.g., \citealt{Berg2021}; \citealt{Olivier2022}). Also, emission lines of different critical densities originate from regions of different densities. Therefore, we further divide the lines of each group in the first step by the ionization potentials. For example, OI, OII, and OIII lines are further categorized into three groups. In the end, according to the species and ionization potentials of the ions, the 128 lines are separated into 14 groups.

\begin{figure*}
    \centering
    \includegraphics[width=\textwidth]{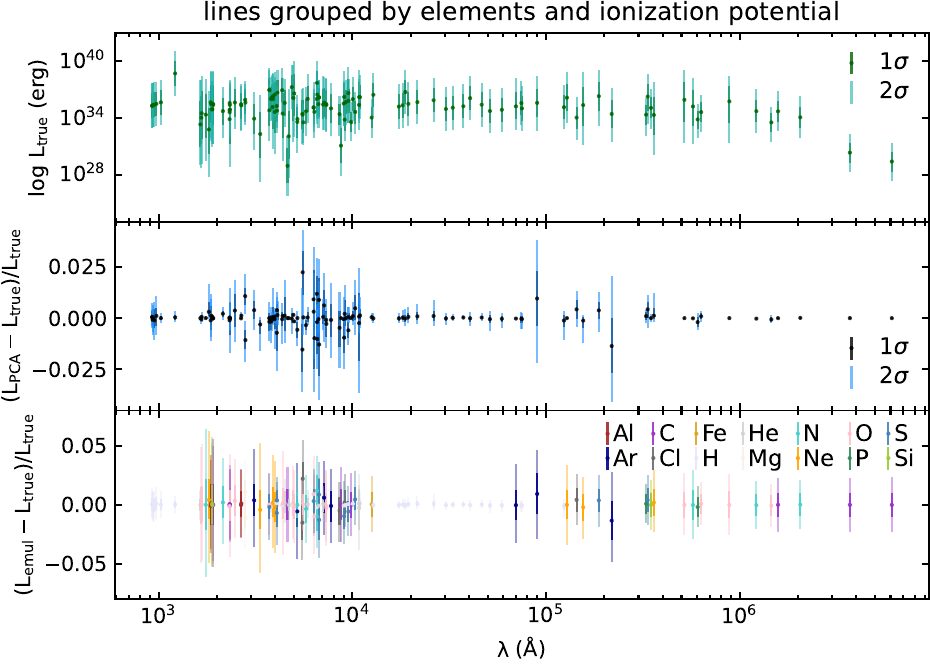}
    \caption{Nebular line emulator errors. The top panel is the distribution of the line luminosities in our training data. The middle panel shows fractional errors of the PCA decomposition using 76 PCA basis. The bottom panel shows the fractional errors of the line emulator for 14 NNs. The element species are denoted by different colors. Note that the colors do not represent our groups of lines for the NNs. We categorize the lines based on both their species and their ionization potentials, and this information is not overplotted in the bottom panel for clarity.} We do not show [Ne\,IV] 4720\,\AA, [Ar\,IV] 7331\,\AA, and  [S\,IV] 10.5\,$\mathrm{\mu}$m here as these high ionization lines exhibit large emulation errors. The rest of the lines have $\lesssim5\%$ errors.
    \label{fig:line_nn}
\end{figure*}

Figure~\ref{fig:line_nn} demonstrates the performance of our line emulator. Across a broad parameter range, the emulator achieves an error at the level of $\lesssim 5\%$ for most lines, accounting for both the adequacy of the PCA basis and the NN accuracy. The hydrogen lines exhibit small errors of $\lesssim 1\%$. Weaker high ionization lines show larger errors. Overall, the emulator uncertainties are sufficiently small compared to the uncertainties in our adopted physics within the photoionization model and observational uncertainties.

Three high ionization state lines, [Ne\,IV] 4720\,\AA, [Ar\,IV] 7331\,\AA, and  [S\,IV] 10.5\,$\mathrm{\mu}$m have 1$\sigma$ emulator errors around 20\%, and are not shown in Figure~\ref{fig:line_nn}. Only high energy photons from $\mathrm{\lambda}<195$\,\AA, $\mathrm{\lambda} < 304$\,\AA, $\mathrm{\lambda} < 356$\,\AA\ can ionize Ne$^{++}$, Ar$^{++}$, and S$^{++}$, respectively. Our power-law parameterizations of a wide range of sources described previously usually do not produce high fluxes at these wavelengths, leading to low number densities of these species. Also, compared to the HeII recombination lines which also probe the high-energy regime, these three collisional excitation lines are usually weaker. Hence, these lines present challenges for the neural network due to their substantial low fluxes in a large fraction of our training data set. We exclude them when performing the mock recovery tests in Section~\ref{subsec:mock}.

\section{Evaluating the performance of the emulator}\label{sec:test}
In this section, we will conduct tests to examine the accuracy of our emulator in reproducing the properties of mock HII regions. In Section~\ref{subsec:SN}, we outline the $S/N$ model that simulates the actual observational uncertainties of the emission lines. Subsequently, we employ this $S/N$ model to generate mock data under the assumption of Gaussian likelihoods, and derive the Fisher information which will provide insights into the lines that the emulator relies more upon. We then fit the mock data with \cue\ and investigate the recovery of the ionizing spectrum and nebular parameters for different sources.

To generate mock data, we take a target from JWST Cycle 1 program Blue Jay (GO 1810; PI Belli) as an exemplar (see the details of the Blue Jay survey and data reduction in Belli et al. in preparation; and the first results of Blue Jay in \citealt{Belli2023}; \citealt{Conroy2023}; \citealt{Davies2023}; \citealt{Park2024}). The Blue Jay survey observed a mass-selected sample ($9 < \log(M_*/M_\odot) < 11.5$) of $\sim$150 galaxies at $1.7 < z < 3.5$ in the COSMOS field. The NIRSpec micro-shutter array was used to obtain $R \sim 1000$ spectra with three medium-resolution gratings (G140M, G235M and G395M) with exposure times of 13h, 3.2h and 1.6h respectively.

\subsection{S/N model}\label{subsec:SN}
We build a simple signal-to-noise model for observed emission lines based on the \prospector{} fit to the photometry and spectrum of a star-forming galaxy from Blue Jay. %target 19572
This $S/N$ model determines the emission line measurement uncertainties given the line luminosities. It will be used later in Section~\ref{subsec:fisher} and Section~\ref{subsec:mock} to generate the mock emission lines. Note that we do not consider the emulator uncertainty for the tests in Section~\ref{subsec:fisher} and Section~\ref{subsec:mock}. The emulator uncertainty for a specific emission line involves complex dependencies on emulator parameters and PCA decomposition uncertainties, and it does not follow a Gaussian distribution. Additionally, as illustrated in Figure~\ref{fig:line_nn}, the emulator uncertainty is generally small, compared to the observational uncertainty, except for high ionization state lines.

We derive the signal and the noise of 128 emission lines using line predictions from the posterior of the \prospector{} fit. The signal vector $\boldsymbol{S} = \{S_{i, 1}, \ldots, S_{i, 128}\}$ is computed from the posterior-weighted sum $\boldsymbol{S} = \sum w_i \boldsymbol{L_i}$, where $w_i$ denotes the weight of the $i$-th \prospector{} posterior sample, and $\boldsymbol{L_i}=\{L_{i, 1}, \ldots, L_{i, 128}\}$ represents the model emission lines in this sample. In our \prospector{} setup, we enable emission line marginalization (see Appendix E of \citet{prospector}).
%which imposes a penalty on the model likelihood. 
In this context, the chosen emission lines are modeled by a Gaussian, and the model uncertainty of emission lines arises from the mean line luminosity uncertainty due to the SED continuum uncertainty and the Gaussian uncertainty around the mean. Therefore, the noise term from the mixture of Gaussian distributions can be expressed as $\boldsymbol{N} = \sqrt{\sum {w_i (\boldsymbol{L_i}^2+\boldsymbol{\sigma_i}^2)} - (\sum {w_i \boldsymbol{L_i}})^2}$, where $\boldsymbol{\sigma_i} = \{\sigma_{i, 1}, \ldots, \sigma_{i, 128}\}$ is the standard deviation of Gaussian fits to the emission lines. The final observed signal-to-noise ratio $\boldsymbol{\xi}$ is then given by 
\begin{equation}
    \boldsymbol{\xi} = \sum_{n} w_i \boldsymbol{L_i} / \sqrt{\sum_{n} {w_i (\boldsymbol{L_i}^2+\boldsymbol{\sigma_i}^2)} - (\sum_{n} {w_i \boldsymbol{L_i}})^2},
\end{equation} 
where n is the posterior sample size. For the emission lines outside the observation wavelength range, we first estimate a $\boldsymbol{S}$--$\boldsymbol{\xi}$ relationship from the lines residing within the Blue Jay wavelength range, and then use this function to estimate the signal-to-noise ratio $\xi$ of those unobserved lines according to their $S$ from \prospector. For our chosen star-forming galaxy from Blue Jay, we obtain a $S/N$ for $\mathrm{H}\alpha$ of 8.53, and the lowest $S/N$ of the 128 emission lines is 0.48. In Section~\ref{subsec:fisher} and \ref{subsec:mock}, we estimate the $S/N$ model at different $S/N (\mathrm{H}_\alpha)$ by scaling this $\boldsymbol{\xi}$ up or down. To be more specific, the $S/N$ of every emission line is multiplied by the same factor $S/N (\mathrm{H}_\alpha) \over 8.53$ to achieve the desired $S/N (\mathrm{H}_\alpha)$.

\subsection{Fisher Information of the line emulator}\label{subsec:fisher}
\begin{figure*}
    \centering
    \includegraphics[width=0.9\textwidth]{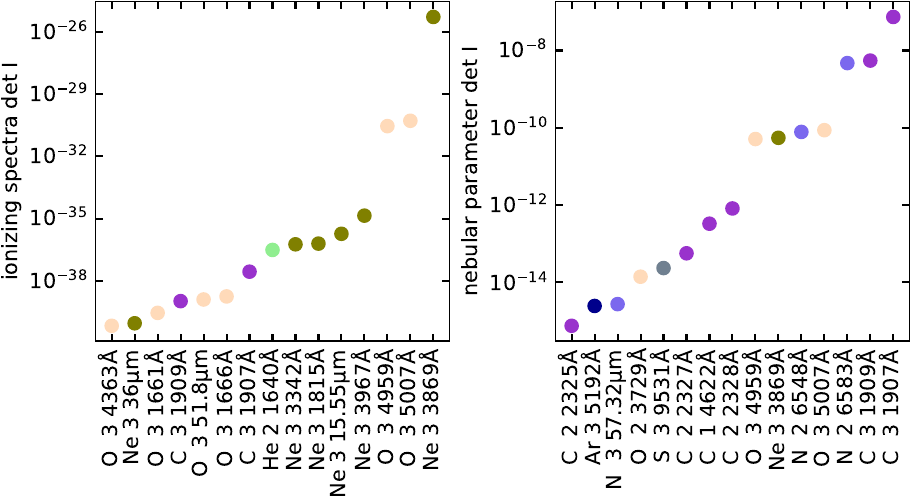}
    \caption{Determinant of the Fisher information matrix for ionizing spectrum parameters and nebular parameters, respectively. Different elements are marked by different colors. Only the 15 lines with highest $\det I_\mathrm{1, ..., 128}(\btheta)$, i.e. that are most important in determining the ionizing or nebular properties, are shown here. These lines exhibit higher sensitivity to parameter changes, making them more crucial for constraining parameters. The O\,III and Ne\,III lines are particularly effective in constraining the ionizing spectrum shape. The C, N, and O lines are most informative for inferring the nebular properties.}
    \label{fig:fisher_all}
\end{figure*}
\begin{figure*}
    \centering
    \includegraphics[width=0.9\textwidth]{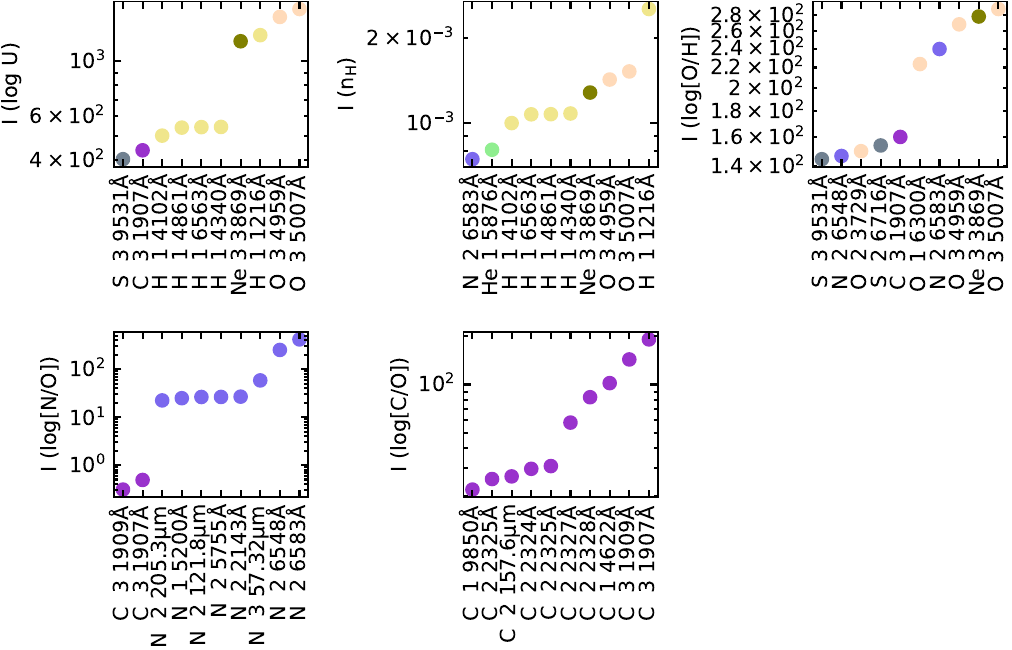}
    \caption{Fisher information for each nebular condition parameter. Different elements are denoted by different colors. We display the 10 lines with the highest Fisher information. As described in detail in Section~\ref{subsec:fisher}, the results align reasonably well with expectations from nebular physics. This can be used to identify important lines for measuring a specific parameter: for example, [OIII] doublets at 5007\,\AA\ and 4959\,\AA\ are important for constraining $\log U$, $\nH$, and [O/H].}
    \label{fig:fisher_neb}
\end{figure*}

We leverage the Fisher information matrix $\boldsymbol{I}(\theta)$ to investigate the sensitivity of the lines to changes in the emulator parameters. This gives an intuitive understanding of, given a set of observed lines, including both flux and error, which lines are important in constraining a given parameter. The Fisher information matrix is defined as the variance of the partial derivative of the log-likelihood function $\ln p(\bX, \btheta)$ with respect to the parameters $\btheta$. 
\begin{equation}
\boldsymbol{I}_{i j}(\btheta)=\int_{\bX} \frac{\partial \ln p(\bX, \btheta)}{\partial \theta_i} \frac{\partial \ln p(\bX, \btheta)}{\partial \theta_j}p(\bX, \btheta) d{\bX}, ~\rm{where}
\end{equation}
\begin{equation}\label{eq:mock}
\ln p(\bX, \btheta) = \sum_{\mathrm{n} = 1}^{128} -\frac {\ln (2 \pi \boldsymbol{\sigma}_\mathrm{mock}^2)}{2} - \frac{{(\bX_{\mathrm{mock}} - \bX_{\mathrm{predicted}})}^2}{2 \boldsymbol{\sigma}_\mathrm{mock}^2}.
\end{equation}
Here $\btheta$ represents the model parameters described in Section~\ref{sec:cloudy}. $\bX = \{x_1, ..., x_{128}\}$ is the mock data, the emission line luminosity predictions from our emulator perturbed by a Gaussian uncertainty given the $S/N$ model. The observation uncertainty is $\boldsymbol{\sigma}_\mathrm{mock} = \bX_{\mathrm{mock}}/\boldsymbol{\xi_{10}}$, where we scale the observed $\boldsymbol{\xi}$ to get the signal-to-noise ratios when $S/N (\mathrm{H}_\alpha) = 10$, i.e., $\boldsymbol{\xi_{10}}$. We compute the determinant of the Fisher information matrix for every line $\det I_\mathrm{1, ..., 128}(\btheta)$, quantifying the amount of information an observable carries about the parameters. 
%The Cramer-Rao theorem states that any unbiased estimator for the parameters will deliver 
The inverse Fisher matrix $I(\boldsymbol{\theta})^{-1}$ provides the lower bound of the covariance matrix on the parameters. As $\boldsymbol{I}(\theta)$ increases, the variance of the estimator decreases, making it easier to measure the parameters. Note that the Fisher information depends on both the $S/N$ model and the emulator uncertainty. Therefore, our results based on the $S/N$ of a normal star-forming galaxy might change for other types of ionizing sources. %The Fisher matrix therefore offers a best-case scenario for ones ability to constrain the model parameters given a set of observations.

In Figure~\ref{fig:fisher_all}, we present the determinant of the Fisher information matrix for the seven ionizing spectrum shape parameters and the five nebular parameters. The [O\,III] and [Ne\,III] lines exhibit the highest sensitivity for determining the ionizing spectrum shape. Apart from being very bright (and thus having high $S/N$) and influencing the likelihood significantly, the [O\,III] lines originate from the ionizing spectrum blueward of 353\,\AA, where different ionizing sources demonstrate the most significant differences. Ne$^{++}$ is iso-electric equivalent to O$^{++}$, and Ne$^{++}$ energy levels are similar to those of O$^{++}$ (e.g., \citealt{Nagao2006}; \citealt{Zeimann2015}). Thus, the [Ne\,III] lines behave similarly as the [O\,III] lines. The carbon, nitrogen, and oxygen lines are the most important for inferring the ionization strength, gas density, and chemical abundances. This result aligns with expectations, as [O/H], [C/O], and [N/O] constitute three of our five nebular parameters.

We further examine the Fisher information for individual nebular parameters in Figure~\ref{fig:fisher_neb}. In this case, the Fisher information simply implies the variance of each parameter. The hydrogen lines, [O\,III] doublet, and [Ne\,III] 3869\AA\ carry the most information about the ionization strength and gas density. This is expected as [O\,III]/HI ratios are sensitive to the ionizing photon budget. Strong metal lines are most important for determining the metallicity represented by [O/H], as the strength of these collisional excitation lines depends on the temperature and the cooling efficiency of the ionizing gas, with oxygen playing a major role as a coolant. The [N/O] ratio is most sensitive to nitrogen lines. Similarly, the [C/O] ratio is most sensitive to carbon lines. The results from Figure~\ref{fig:fisher_all} and Figure~\ref{fig:fisher_neb} suggest that our neural net emulator is sensible of the well-known connections between nebular conditions and observed lines while training.

\subsection{Mock tests}\label{subsec:mock}
Our emulator is affected by three sources of uncertainty, the power-law approximation, the PCA decomposition, and the NN training. We conduct mock tests for SSPs, post-AGBs, and AGNs to assess the emulator performance when all three uncertainties are taken into account at $S/N (\mathrm{H}_\alpha) = 1, 10, 100$. For each ionizing source type and each $S/N (\mathrm{H}_\alpha)$, we generate 1,000 mock emission line observations using the $S/N$ model described in Section~\ref{subsec:SN}. We use the spectra of SSPs, post-AGBs, and AGNs (Figure~\ref{fig:ion_spec}) as ionizing spectra, and randomly draw the nebular parameters within their training range (Table~\ref{tab:cloudy_prior}). True emission line luminosities are obtained by running \cloudy, and we introduce perturbations based on their uncertainties at the given $S/N (\mathrm{H}_\alpha)$ to simulate mock observations. According to the wavelength coverage of the Blue Jay data, our mock tests cover all emission lines between 1250\AA\ and 12000\AA. We use the same uniform priors as in the training set (Table~\ref{tab:cloudy_prior}), and the likelihood is defined by Equation (\ref{eq:mock}). The posterior parameter distribution is sampled using the dynamic nested sampling code \texttt{dynesty} (\citealt{Speagle2020}; \citealt{Koposov2022}).

\begin{figure*}
    \centering
    \includegraphics[width=\textwidth]{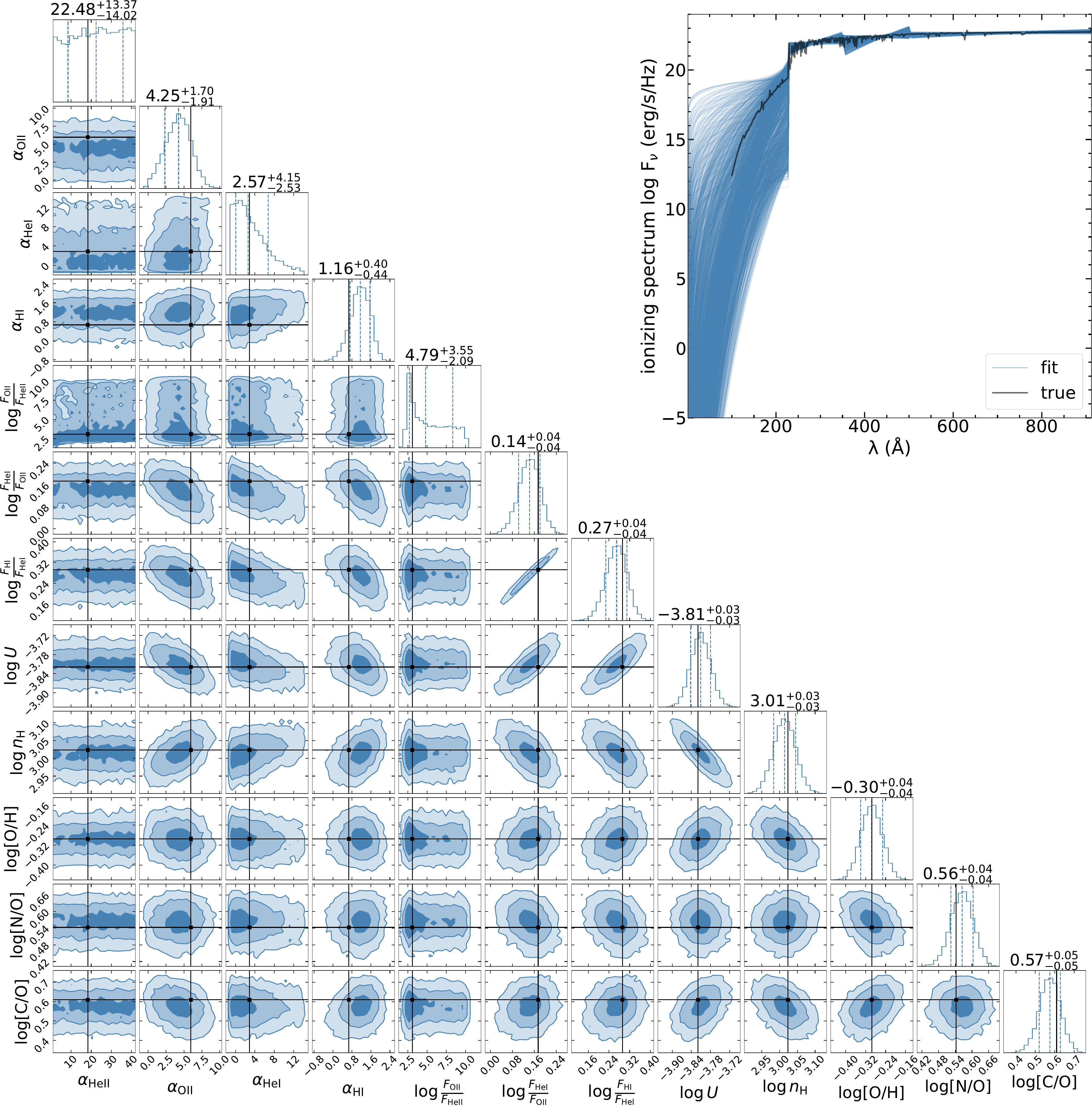}
    \caption{Mock recovery test for an object ionized by an SSP at an age of 2.2\,Myr and a stellar metallicity $\log (Z/Z_\odot) = -1.2$, assuming $S/N (\mathrm{H}_\alpha) = 10$. The contours show the 1$\sigma$, 2$\sigma$, 3$\sigma$ of the posterior. The dashed lines mark the 16th, 50th, and 84th percentiles. The truths are in black. We also demonstrate in the upper right the true ionizing spectrum and 2,000 random draws from the posterior. \cue\ is able to recover the true input parameters within 2$\sigma$ of the posterior.}
    \label{fig:mock_one}
\end{figure*}

\begin{figure*}
    \centering
    \gridline{\fig{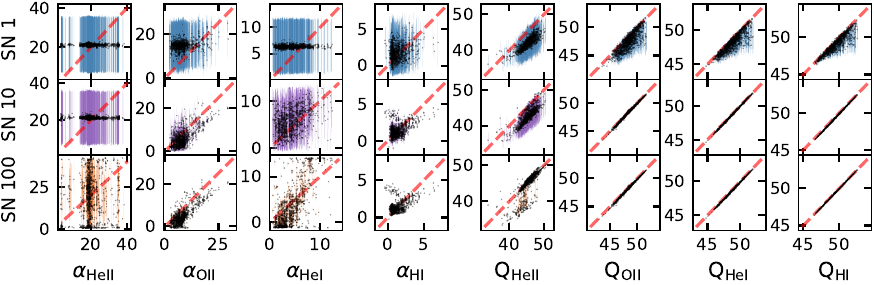}{\textwidth}{(a) ionizing spectrum.}}
    %\hspace*{\fill}
    \gridline{\fig{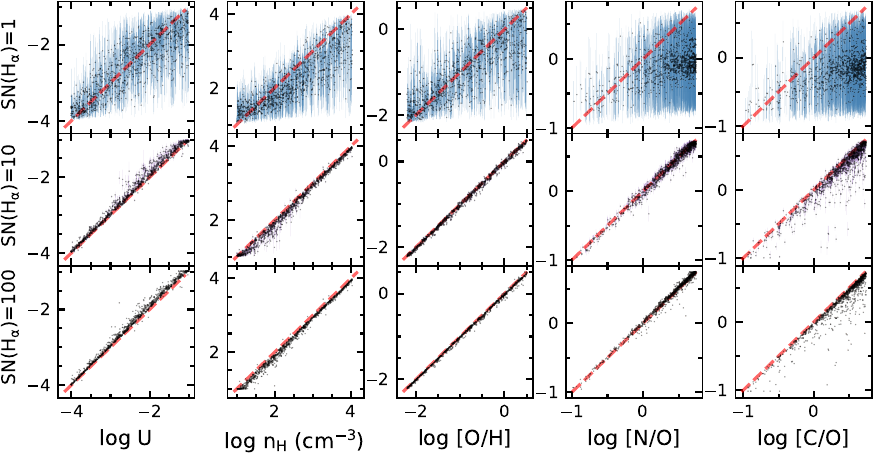}{\textwidth}{(b) nebular properties.}}
    \caption{Mock recovery tests for objects with stellar type ionizing spectrum. From the top to the bottom row: $S/N$ (H$_\alpha$) $= 1, 10, 100$. The x-axes are the input parameters and the y-axes are the recovered parameters from our emission line emulator. The posterior median is in black. The error bars are the 1$\sigma$ range of the posterior. The red dashed line marks the 1:1 line. The truths for the nebular parameters lie within $\sim$0.2\,dex of the posterior median at $S/N (\mathrm{H}_\alpha) \gtrsim 10$. For the ionizing spectrum parameters, we are not able to recover the power-law indexes, but we achieve a good estimate of the number of ionizing photons emitted at each segment of the ionizing spectrum, which is the main determination of the ionizing spectrum shape.}
    \label{fig:mock_ssp}
\end{figure*}

Figure~\ref{fig:mock_one} illustrates the parameter recovery test for one mock observation at $S/N (\mathrm{H}_\alpha) = 10$. Both the ionizing spectrum parameters and the five nebular parameters are recovered within 1$\sigma$. Note that the typical offsets of the power-law parameters are larger as we show later in Figure~\ref{fig:mock_ssp} The extreme-UV part of the ionizing spectrum has a wide posterior. Specifically, the posterior of the power-law parameters at the bluest bin is nearly flat. This part of the ionizing spectrum usually contributes fewer ionizing photons and is probed by weak lines, resulting in a less constrained posterior, which is nonetheless well-calibrated (i.e., the posterior correctly reflects our lack of constraints). Furthermore, the contours of the posterior reveal several correlations among the inferred parameters. The flux ratios of the $228\mathrm{\AA}<\lambda<353\mathrm{\AA}$, $353\mathrm{\AA}<\lambda<504\mathrm{\AA}$, and $504\mathrm{\AA}<\lambda<912\mathrm{\AA}$ segments are correlated, in good agreement with no strong break at the HeI and OI ionization edge in this SSP. The ionization strength and gas density display an anti-correlation, reflecting their similar effects on the total ionizing budget. The elemental abundances [O/H], [C/O], and [N/O] show marginal covariance as they are all sensitive to the nebular temperature and density structure. The rest of the parameters do not exhibit strong covariances.

Figure~\ref{fig:mock_ssp} presents the mock test for SSPs for different $S/N$. In Figure~\ref{fig:mock_ssp}(a), the ionizing photon rate at each power-law segment primarily determines the shape of the ionizing spectrum, with the power-law index having more of a secondary effect.
%We effectively capture the first-order shape of the ionizing spectrum through the number of ionizing photon rates at each power-law segment.
As the $S/N$ increases, a highly accurate estimate of total ionizing photons is achieved at each bin, except for a slight underestimation of HeII ionizing photons for $\sim$60\% of the mock samples at $S/N$(H$\alpha$) = 100. Comparatively, unbiased estimates of HeII ionizing photons and $\alpha_\mathrm{HeII}$ are obtained for AGN ionizing spectra at $S/N (\mathrm{H}_\alpha) = 100$ (Figure~\ref{fig:mock_agn}), where these ionizing spectra have higher $Q_\mathrm{HeII}$. 
This suggests that the emulator provides a lower bound of the extreme-UV photons if we do not observe strong emission lines powered by these photons. Even if the emulator does not provide useful constraints for the HeII ionizing photons based on a given set of observed emission lines, the ratios of ionizing photons between the redder piece-wise regions may still provide enough information to distinguish between stellar populations, post-AGBs, and AGNs as shown in Figure~\ref{fig:ion_spec} and the mock tests.

Although our emulator effectively recovers the first-order shape of the ionizing spectrum through ionizing photon rates at each power-law segment, it has some difficulties in inferring the detailed slope. At $S/N (\mathrm{H}_\alpha) = 1$, the power-law indexes are not constrained with their inferred posterior median close to the prior median and their posterior width comparable to the prior range. The fact that the posterior of power-law indexes effectively follows the prior indicates that at $S/N (\mathrm{H}_\alpha) = 1$, the emulator does not have sufficient information to constrain the power-law index. At $S/N (\mathrm{H}_\alpha) > 10$, \cue\ still cannot recover $\alpha_\mathrm{HeII}$ for the SSPs (Figure~\ref{fig:mock_ssp}) because the emission lines ionized by the HeII ionizing spectrum are typically very weak for stellar populations. It implies that \cue\ cannot infer the power-law indexes when no strong emission lines ionized by the corresponding power-law piece are detected. However, \cue\ is able to recover $\alpha_\mathrm{HI}$ in most cases as observed in Figure~\ref{fig:mock_ssp}, Figure~\ref{fig:mock_pagb}, and Figure~\ref{fig:mock_agn}. Moreover, it provides constraints on $\alpha_\mathrm{OII}$ for SSPs and post-AGBs, and on $\alpha_\mathrm{HeII}$ for AGNs. These suggest that the best practice of using \cue\ to infer the power-law index is when the observations show strong emission lines ionized by photons from that power-law segment. Note that in practical SED fitting of star-forming galaxies, the light from the synthetic CSPs is dominated by young stars, whose ionizing spectra usually have fewer features and are easier to describe than old stars. Therefore, our emulator is expected to perform better at unveiling the ionizing spectrum in real cases.

Figure~\ref{fig:mock_ssp} also reveals our ability to constrain nebular properties at $S/N (\mathrm{H}_\alpha) \geq 10$. At $S/N (\mathrm{H}_\alpha) = 1$, the wide posterior encompasses the input nebular parameters within 1$\sigma$. At higher $S/N$, the offsets between the input and the inferred parameters are generally $<0.2$\,dex for all nebular parameters. At $S/N (\mathrm{H}_\alpha) = 100$, the uncertainty of the estimates is small and primarily dominated by the emulator error. In addition to the observational uncertainty, the wavelength coverage may influence the efficacy of mock tests. As indicated by the Fisher information, lines carrying the most information are predominantly optical. The absence of UV and NIR emission lines could compromise constraints on [C/O] and the ionizing spectrum.

The inferred [C/O] shows a slight bias even at $S/N (\mathrm{H}_\alpha) \geq 10$. As discussed in Figure~\ref{fig:mock_one}, [C/O] exhibits strong covariance with $U$, $\nH$, and [O/H], affecting the accuracy of [C/O] estimates due to interactions with other nebular parameters. We find that the offsets between the input and inferred [C/O] are correlated with the contrast of the ionizing spectrum to the UV--optical spectrum.  Lowering $R_\mathrm{ionizing}$ results in larger errors in [C/O] estimation. This pairs up with the fact that the inferred [C/O] is not biased for the post-AGB ionizing spectrum and AGN ionizing spectrum (see Figure~\ref{fig:mock_pagb} and Figure~\ref{fig:mock_agn}), because they have greater $R_\mathrm{ionizing}$ and their UV-optical photons contribute less to illuminating the HII region.
%For $U$ and $\nH$, their inferred values are mostly consistent with the truth. The bias might be due to sampling and the emulator uncertainty. 

In Appendix~\ref{sec:appendixA}, we further show our mock test results for the AGN and post-AGB powered nebular emission. The conclusions for these tests are broadly consistent with the SSP mock tests. The nebular parameters and the total ionizing radiation at each power-law bin are overall successfully recovered within the uncertainty, while the power-law indexes are not well constrained.

\section{Discussion} \label{sec:discussion}
\subsection{An outlook: applications of \cue\ in a wide range of environments}
\cue\ has broad applications, offering a versatile tool for rapidly modeling nebular line and continuum emission in a wide range of environments and conducting rapid population studies of nebular properties and ionizing sources. Its adaptability to various ionizing sources and stellar models allows for easy integration into a diverse range of nebular properties. As a showcase, we apply \cue\ to modeling a peculiar nebular galaxy at $z=5.943$ and interpret the inferred ionizing spectrum with a mixture of stars and AGN in \citet{Li2024}.

Given its capability of directly inferring the ionizing spectrum from emission lines, one natural application of \cue\ is to interrogate the stellar models. We perform a mock recovery test to evaluate the discriminative power of \cue\ between BPASS and Padova stellar models. The impact of binary interactions on the ionizing spectrum and emission lines interpretations has been established in literature (e.g., \citealt{Steidel2016}, \citealt{Gotberg2017}, \citealt{Stanway&Eldridge2019}, \citealt{Gotberg2019}, \citealt{Ma2022}, \citealt{Eldridge2022}). The binary evolution model in BPASS prolongs the ionizing output from massive stars, and powers a harder ionizing spectrum with more ionizing photons compared to the single-star Padova model. Following the procedures in Section~\ref{subsec:mock}, we generate mock emission lines from the same SED model of a star-forming galaxy using the BPASS, and Padova isochrones separately with the MILES library and the \citet{Chabrier2003} IMF. We then use \cue\ to infer the ionizing spectrum from the mock emission lines. This tests if we can differentiate the ionizing CSPs from the two stellar models conditioned on the case that we can sufficiently narrow down the ionizing CSPs predicted by the stellar population synthesis model. This requires the population synthesis model to have tight constraints on the other SED model parameters including SFH, stellar metallicity, dust properties, IMF, etc. 
%In the actual SED fitting, our posterior of the ionizing spectrum may be wider than this ideal case.

\begin{figure*}
    \centering
    \includegraphics[width=0.85\textwidth]{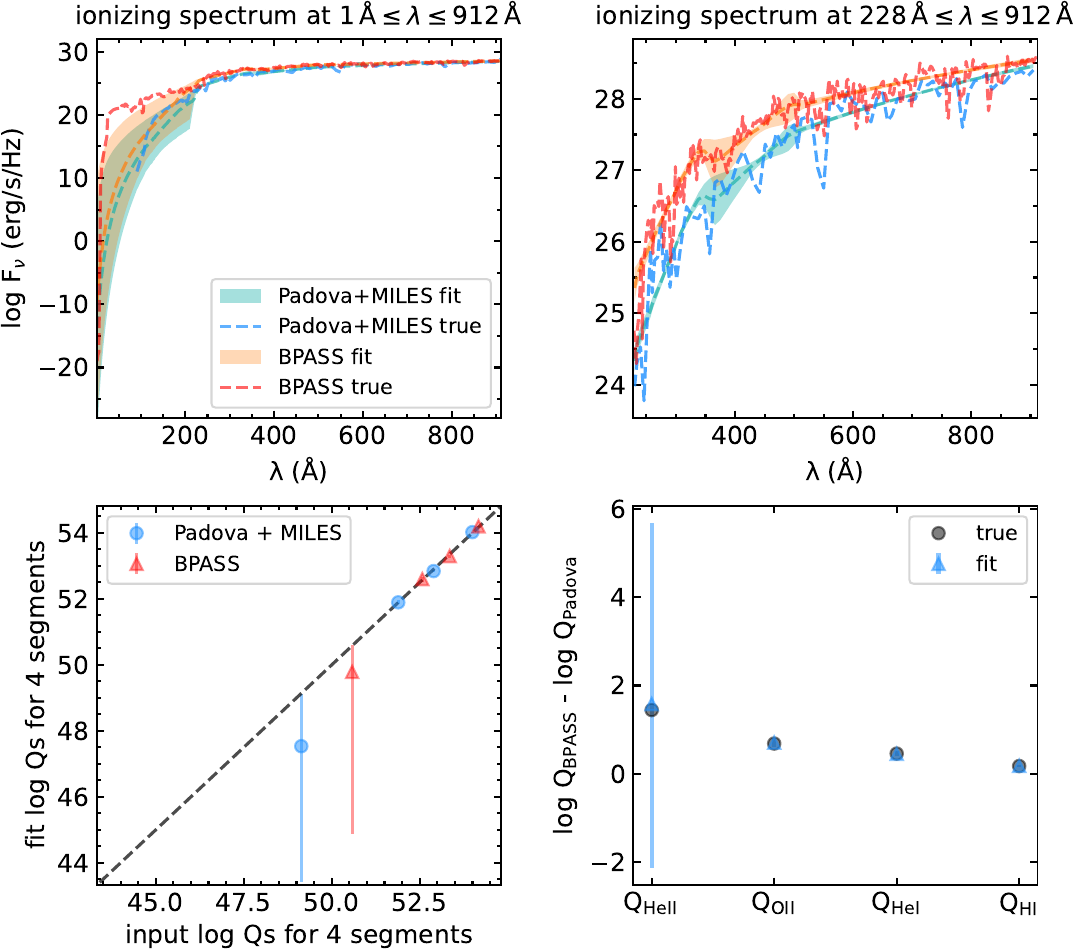}
    \caption{The recovery tests on the single-star model Padova and the binary evolution model BPASS. Top left: input and recovered ionizing spectra of a star-forming galaxy. Top right: a zoom-in view of the red part of the ionizing spectra. Bottom left: the input and recovered ionizing photon input at each power-law segment. Bottom right: the difference between the two stellar models regarding the ionizing photon input. Our emulator can distinguish between Padova and BPASS stellar models at the redder part of the ionizing spectrum.}
    \label{fig:padova-bpass}
\end{figure*}

The analysis results are shown in Figure~\ref{fig:padova-bpass}. The ionizing spectrum of the tested BPASS CSP has more ionizing flux than the Padova CSP at all wavelengths. In particular, BPASS ionizing spectrum produces $>$10 times more HeII ionizing flux. %, although our emulator struggles to tightly constrain the hardness of the HeII ionizing spectrum.
The recovered ionizing spectra are consistent with the input ionizing spectrum at the $2\sigma$ level, with a tendency to underpredict HeII ionizing photons for both stellar models. The models deviate most in the hard ionizing spectrum. However, this turns out to not be the best way to distinguish them, because the wide posterior of the HeII ionizing spectrum washes out their difference (see the lower right panel of Figure~\ref{fig:padova-bpass}). 

Instead, our emulator is very accurate at revealing the number of ionizing photons at the three redder segments of the ionizing spectrum. The input BPASS ionizing spectrum has $\QH = 1.74 \times 10^{54}$\,s$^{-1}$, and the input Padova ionizing spectrum has $\QH = 1.08 \times 10^{54}$\,s$^{-1}$. \cue\ is able to retrieve the true $\QH$ within 5\%. Furthermore, in the upper right panel of Figure~\ref{fig:padova-bpass}, we demonstrate that \cue\ is capable of revealing the distinct shape and the normalization of the ionizing spectrum of BPASS and Padova at $\lambda \gtrsim 200$\,\AA. So, we can draw a conclusion that even though our emulator cannot put tight constraints on the HeII ionizing spectrum, it is sufficient to differentiate the BPASS and Padova model by capturing their differences in the ionizing spectrum shapes and normalizations at the redder part of the ionizing spectrum. Since binary stellar populations also play a critical role in the stellar feedback and chemical enrichment of galaxies (e.g., \citealt{Kobayashi2020}, \citealt{Doughty2021}, \citealt{Eldridge2022}, \citealt{Yates2023}), the nebular conditions around the stars may also differ for these two stellar models, which may further help \cue\ distinguish the models. 

While we have demonstrated that \cue\ has the potential to distinguish between single and binary stellar populations based on their ionizing shape at $\lambda \gtrsim 200$,\AA, there are complexities involved in applying this to real observations. One important uncertainty is the dust attenuation of the ionizing photons before they reach the gas cloud. Because the inferred ionizing radiation by \cue\ represents the incident flux striking the inner face of the gas cloud, it becomes challenging to trace back to the source ionizing spectrum when dust attenuation is significant. For example, \citet{Tacchella2022} shows that dust attenuation within the HII region can not only reduce the number of ionizing photons by $\sim$30\% for Milky Way-like galaxies but also lead to a harder ionizing spectrum as it affects ionizing photons from younger stars more than old stars. In addition, as mentioned earlier, this test is conditioned on the perfect knowledge of the CSP properties like SFH and stellar metallicity. In future work we will explore how well \cue\ can distinguish between stellar models in real observations when the ionizing spectrum shape has some degeneracy with other model parameters.

The emulator's speed allows for industrial-scale investigations into nebular properties and ionizing source natures. We can model spatially resolved HII regions (e.g., the CHemical Abundances Of Spirals (CHAOS) project \citep{Berg2020}; the JWST TEMPLATES Early Release Science program) or large-scale spectroscopic surveys (e.g., the Sloan Digital Sky Survey (SDSS; \citealt{Abdurrouf2022}); the Prime Focus Spectrograph survey (PFS; \citealt{Greene2022}). These could reveal, for example, the resolved ionizing properties of various galaxy populations, and the connections between stellar populations and ionizing gas. Because \cue\ can describe the ionization by hot evolved stars and black holes too, its applications are not limited to the star-forming regions, with the caveat that \cue\ is not trained to model the ionization by old main sequence stars.

\subsection{Comparison to previous work on interpreting emission line ratios}

\cue\ expands the model parameter space especially for the ionizing spectrum compared to previous nebular emission studies based on grids of photoionization models (e.g., \citealt{Byler2017}, \citealt{Gutkin2016}, \citealt{Feltre2016}). Like these tools, we can use \cue\ inside of a galaxy SED-fitting code like \prospector. 

\begin{figure}
    \centering
    \includegraphics[width=0.85\columnwidth]{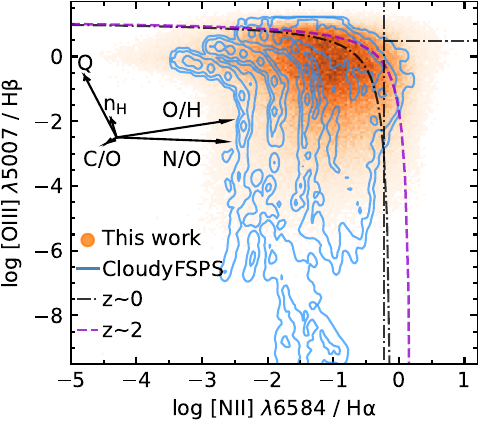}
    \caption{The location of \cue\ training data and \citet{Byler2017} model grids on the BPT diagram. The black lines demarcate AGNs and star-forming galaxies, and split Seyfert galaxies and LINERs in the AGN regime at $z \sim 0$ \citep{Kauffmann2003}. The purple line shows the measured locus of the $z \sim 2.3$ sample \citep{Strom2017}. The arrows annotate the average moving direction when changing the parameter.}
    \label{fig:BPT}
\end{figure}

In Figure~\ref{fig:BPT}, we show \cue's coverage in the traditional BPT diagram. Compared to \citet{Byler2017} (labeled as CloudyFSPS which our \cloudy\ settings reference from), \cue\ made several updates: it is able to model a wide range of ionizing sources by describing the ionizing radiation with a general parametric form; the stellar and gas phase metallicity are uncoupled; gas density, [N/O], and [C/O] are modeled as free parameters. These changes allow \cue\ to explain the line ratios beyond the typical star-forming galaxy regime. Particularly, Figure~\ref{fig:BPT} shows that \cue\ can describe emission line ratios resembling AGNs, which typically occupy the upper right region of the BPT diagram, even though our training set includes fewer AGNs than stellar ionizing sources.

Nebular conditions and ionizing sources evolve with redshift, causing the line ratio diagrams to evolve with redshift (\citealt{Steidel2014}; \citealt{Strom2017}; \citealt{Garg2022}; \citealt{Hirschmann2023a}; \citealt{Hirschmann2023b}). For instance, explanations for the redshift evolution of the BPT diagram often include a higher ionization parameter, a harder ionizing radiation field driven by $\alpha$-enhancement, or an increasing N/O towards higher redshifts. \cue\ can interpret this evolution by turning line evolution into redshift evolution of parameters and properties. Figure~\ref{fig:BPT} illustrates the degeneracy between the nebular conditions and the ionizing source captured by \cue. We show here the average response of the line ratios to the change of the nebular parameters. A higher metallicity, and a larger [N/O] drive a higher [NII] 6564\AA/ H$\alpha$ ratio. More ionization photon input drives a higher [OIII] 5007\AA/ H$\beta$ ratio. These effects can move star-forming galaxies to the AGN regime on the BPT diagram. This manifests the need for a more flexible nebular model like \cue\ to interpret the emission line ratios self-consistently. 

\subsection{Limitations, caveats, and areas for future improvement}
Our emulator's efficacy is subject to the complexity of nebular emission modeling, including geometry, abundance sets and dust depletion factors, the approximation of seeing the whole galaxy as one HII region, and the unmodeled emission from diffuse ionized gas (DIG). While Cue offers flexibility to model different sources, there are additional sources of ionization that \cue\ may be less effective at describing, such as shocks and massive X-ray binaries.

Observations indicate the importance of shock ionization for starburst galaxies and AGN outflows (e.g., \citealt{Soto2012}; \citealt{Kewley2013}). However, it is challenging to identify the contribution of shocks to the nebular emission. The shock models overlap with the AGNs and star-forming galaxies on the BPT diagram, with similar trends between the line ratios and the nebular parameters as AGN models (e.g., \citealt{Hirschmann2023b}). %Potentially, we could use [FeII]1.257 to identify the presence of shocks \citep{Brinchmann2023}.
Moreover, in the real universe, shocks usually are concurrent with stellar populations or AGNs to ionize the surrounding gas. Therefore, to model shock ionization properly, we need to locate the shocks using spatially resolved data and separate out their contribution to the emission lines using the kinematic information (e.g., \citealt{Duncan2023}; \citealt{Johnston2023}). In addition to the observational challenge, most studies on the shock and precursor ionization model utilize the pressure-based photoionization code MAPPINGS (e.g., \citealt{Allen2008}; \citealt{Rich2010}; \citealt{Alarie&Morisset2019}). Compared to our constant density assumption, such constant pressure assumption is more reasonable for the shock-ionized region. Our \cloudy\ emulator based on uniformly distributed gas may have difficulty reproducing the strong coronal lines of Seyfert galaxies (e.g., \citealt{Zhu2023}). The non-photoionization and non-collisional equilibrium nature of shocks also makes them difficult to model. Given these complexities, modeling ionization and subsequent reemission from shocks is beyond the scope of this work. 

Aside from shocks, X-ray binaries are also a compelling source of high-ionization emission lines (e.g., \citealt{Senchyna2020}, \citealt{Garofali2023}), which \cue\ has large uncertainties on. The ionizing spectra shapes of X-ray binaries are relatively simple and can be parameterized by power-laws. It is straightforward to extend the wavelength and ionizing spectrum coverage of \cue\ to the regime of X-ray binaries and this may be the subject of future work.

In addition to HII regions, DIG can also contribute to the integrated emission lines on large scales. Emission from the DIG is characterized by a harder ionizing spectrum, lower ionization parameter, lower density, and lower pressure compared to the HII region (e.g., \citealt{McClymont2024}). It makes up $\lesssim$60\% of the integrated high ionization lines and hydrogen lines but may contaminate the observed low ionization lines from HII regions by up to $\sim$200\% (e.g., \citealt{Blanc2009}; \citealt{Kreckel2016}; \citealt{Zhang2017}; \citealt{Tacchella2022}), especially [SII] 6717\,\AA\ and [SII] 6731\,\AA. DIG can be excited by the leaked ionizing radiation from the HII region, post-AGBs, or shocks. Thus, the DIG poses challenges in reconciling with simple photoionization models. The contribution of DIG to the integrated nebular emission of galaxies needs to be appropriately modeled when fitting photometry and spectroscopy with fixed-size apertures (e.g., \citealt{Mannucci2021}).

Apart from the grid-based nebular model, 3D hydrodynamical simulations are another approach to understanding galaxy nebular emission. This can be done via on-the-fly radiation hydrodynamics on the gas particles (e.g., \citealt{Katz2023a}) or in post-processing of numerical simulations (e.g., \citealt{Smith2022}; \citealt{Tacchella2022}). Compared to 1D photoionization codes like \cloudy, these studies use a more realistic 3D geometry (e.g., \citealt{Haid2018}) for the multiphase interstellar medium and moreover, some of them can take non-equilibrium cooling into account. But owing to the computational cost of the simulations, it is difficult to run enough simulations and achieve dense sampling to compress their nebular predictions to a grid. Hence, it is challenging to train a nebular model between the simulation parameters and the nebular emission. There have also been increasing efforts in combining grid-based nebular models with cosmological simulations built from first principles to understand the nebular line ratios and feedback processes in galaxies across cosmic time (e.g., \citealt{Hirschmann2023b}). These studies can be particularly useful in finding diagnostics of different ionizing sources in the early universe where the number density and spectral feature of each type of source are unconstrained. However, there are some limitations in nebular emission studies with simulations. Different simulations have different behaviors in the star formation history and chemical evolution of galaxies, and use different models for the ionizing sources themselves, making it not straightforward to reconcile the nebular emission predictions among different simulations and between simulations and observations. In addition, the finite spatial and mass resolution of simulations may lead to inaccurate line predictions when the Stromgren radius is not resolved.

One important approximation of \cue\ is modeling the ionizing gas as one HII region. The integrated nebular emission of galaxies is in fact the light-weighted average of multiple HII regions, making the nebular parameters of \cue\ the effective parameters. Resolved maps of emission lines show that the gas and dust properties are not distributed uniformly across the galaxies (e.g., \citealt{Lagos2022}; \citealt{Oey2023}). Processes like radiation pressure and stellar winds will alter the gas cloud structure. This is not captured by the spherical geometry and 1D photoionization we assume (e.g., see the comparison in \citealt{Katz2023b}). Spherical geometry is in general a reasonable assumption as ionization parameter mapping of HII regions suggests that they are mostly regular circular objects \citep{Pellegrini2012}, and the 1D models are successful in reproducing emission line ratios from nearby HII regions (e.g., \citealt{Pellegrini2020}). But these assumptions might fail in certain cases, in which situation the effective parameters are no longer adequate in representing the entire gas cloud. For example, it has been suggested that for dusty HII regions, radiation pressure compresses the gas and dust into a shell, suppressing the absorption of ionizing photons and causing the effective density to diverge from the mean density \citep{Draine2011}. Recent numerical simulations of HII regions have included 3D geometry effects (e.g., \citealt{Haid2018}), and optimally we can test the impacts of these model assumptions by comparing the results from numerical simulations to observations.

%discussiong on the dust-to-gas ratio

\subsection{Emulator speed}
Our emulator includes 15 NNs in total for predicting the nebular continuum and emission lines. One prediction takes approximately 6\,ms execution time using an Intel i7 CPU. This is a $\sim 10^4$ times speedup compared to a \cloudy\ run.
%GPUS are 2-3 times faster (NVIDIA A100-PCIE-40GB)

Apart from using \cue\ as a standalone tool to fit emission line fluxes, we can integrate it into SED fitting codes to fit the observed galaxy spectrum consisting of both the stellar and gas emission, and infer the stellar population properties and ionizing gas properties simultaneously. The emulator prediction based on the integrated ionizing spectrum takes $\lesssim 20$\% execution time of a typical likelihood call in galaxy SED fitting (e.g., \citealt{Leja2019b}), which is a speed-up with respect to using the look-up table on SSPs, especially considering the emulator's great freedom in the parameter space. As a side note, if the users prefer to take the CSPs fixed by the stellar population synthesis model as the ionizing spectrum, the most expensive part is fitting the CSPs with the power-laws, which can take $\sim$10--30\,ms. In this case, most of the runtime is devoted to minimizing our customized loss function defined in Equation (\ref{eq:loss}).

%\subsection{Fit an outlier ionizing source}
%We use \texttt{Prospector} \citep{prospector}, a Bayesian galaxy SED-fitting code to simultaneously fit the global host photometry and spectroscopy. We adopt the MIST isochrones \citep{Choi2016}, and the C3K stellar spectral libraries in the Flexible Stellar Population Synthesis (FSPS; \citealt{Conroy2009, Conroy&Gunn2010}) framework. The stellar population is described by redshift, stellar mass, velocity dispersion, stellar metallicity, and a step function nonparametric star formation history (SFH) with 14 time bins. The nebular emission is parameterized by gas-phase metallicity and ionization parameter using the \citet{Byler2017} grid. We assume a flexible two-component dust attenuation model accounting for birth cloud and diffuse dust separately \citep{Kriek&Conroy2013}. Variation in the shape of the attenuation curve is enabled using a parameter dust index \citep{Noll2009}. We also incorporate the contribution of dust emission to the infrared photometry using a three-parameter model \citep{Draine&Li2007}. To fit the spectroscopy and multi-wavelength photometry together, we distort the spectrum with a polynomial to match the spectra shape of the stellar population. We also include a jitter parameter that inflates the spectroscopy uncertainties to account for imperfect JWST flux calibration and slit losses. Finally, we adopt a parameter to describe the fraction of outlier spectral pixels. In summary, the SED model has 28 free parameters.

\section{Summary}
We introduce \cue, a flexible tool for modeling the continuum and line emission from individual HII regions. By leveraging neural networks to emulate the spectral synthetic code \cloudy, \cue\ covers an extensive parameter space, making it suitable for addressing challenges posed by the unique chemistry and ionizing properties of galaxies in the early universe. It models the input ionizing spectra as a flexible 4-part piecewise-continuous power-law, along with freedom in gas density, total ionizing photon budget, [O/H], [C/O], and [N/O]. 

One main feature of \cue\ is its flexibility. Unlike typical nebular models used in galaxy SED-fitting, \cue\ no longer relies on a set of stellar isochrones and spectral libraries, but instead can take in piece-wise power-law ionizing spectra of a wide range and give results in milliseconds, enabling self-consistent nebular emission predictions for specific stellar models such as one with a top-heavy IMF. By approximating the ionizing spectrum in a parametric form, \cue\ allows direct investigation of the ionizing source through emission lines by marginalizing over the incident ionizing radiation and the chemical conditions.

We demonstrate in the paper that the emulator uncertainties are $\lesssim$5\% for both nebular continuum and emission lines, with the power-law approximation introducing an additional $\lesssim$1\textperthousand{} error for the nebular continuum emission and an $\lesssim$5\% error for the emission lines. Our mock tests suggest that \cue\ accurately capture the shape of the ionizing spectra and the nebular properties based on UV--NIR emission lines at $S/N (\mathrm{H}_\alpha) \gtrsim 10$. The fast execution time of $\lesssim$6\,ms per prediction further enhances \cue's appeal for extensive applications in population studies of nebular emission and ionizing source properties across a broad redshift range.

This fast and flexible emulator paves a way for probing the ionizing spectrum of galaxies, which due to neutral gas absorption is only accessible by the nebular emission they power, and thereby interrogating the ionizing source directly. We have demonstrated that the \cue\ can distinguish stellar models by their differences in the ionizing spectrum. Going forward, \cue\ will be a powerful tool in several contexts -- due to its flexibility it can interpret unique sources powered by unusual or mixed ionizing spectra, while also due to its speed it can be applied to industrial-scale spectroscopic surveys. These studies provide insights on the evolution of nebular conditions and ionization sources of galaxies over cosmic times.

\begin{acknowledgments}
Based on observations with the NASA/ESA/CSA James Webb Space Telescope obtained at the Space Telescope Science Institute, which is operated by the Association of Universities for Research in Astronomy, Incorporated, under NASA contract NAS503127. Y.L. and J.L. are supported under Program number JWST-GO-01810.004-A provided through a grant from the STScI under NASA contract NAS503127. SB is supported by the ERC Starting Grant “Red Cardinal”, GA 101076080. RLD is supported by the Australian Research Council Centre of Excellence for All Sky Astrophysics in 3 Dimensions (ASTRO 3D), through project number CE170100013.
\end{acknowledgments}

\begin{appendix}

\section{Mock Test for Post-AGB and AGN-type ionizing spectra} \label{sec:appendixA}
Here we show the results of the mock recovery tests for objects of post-AGB and AGN-type ionizing spectra in Figure~\ref{fig:mock_pagb} and Figure~\ref{fig:mock_agn}, respectively. The process for generating the mock emission lines is the same as the one in Figure~\ref{fig:mock_ssp}. The post-AGB and AGN ionizing spectra are harder and span a narrower range compared to the SSPs. As a result, \cue\ shows a better performance at recovering their ionizing spectrum and nebular properties. Also, the mock results implicate the same conclusion regarding $S/N$, that at $S/N (\mathrm{H}_\alpha) \gtrsim 10$, \cue\ returns well-calibrated values for the inferred nebular properties and ionizing spectrum. As argued in Section~\ref{subsec:mock}, constraining the power-law indexes can be challenging as they are second-order effects compared to the normalization of the power-laws. When there is a lack of detected strong lines powered by ionizing photons from a particular piece of the ionizing spectrum, we lack sufficient information to constrain the index. Therefore, in the case of AGNs, even if the input ionizing spectrum is a simple power-law, the emulator can only provide useful constraints on $\alpha_\mathrm{HeII}$ and $\alpha_\mathrm{HI}$, but not the other two power-law indexes. Additionally, it is important to note that since the adopted $S/N$ model is based on a normal star-forming galaxy, the mock tests for post-AGB and AGN type ionizing spectra are likely to show better recoveries if we use the $S/N$ model of a post-AGB or an AGN observation accordingly.

\renewcommand{\thefigure}{A\arabic{figure}}
\setcounter{figure}{0}

\begin{figure*}
    \centering
    \gridline{\fig{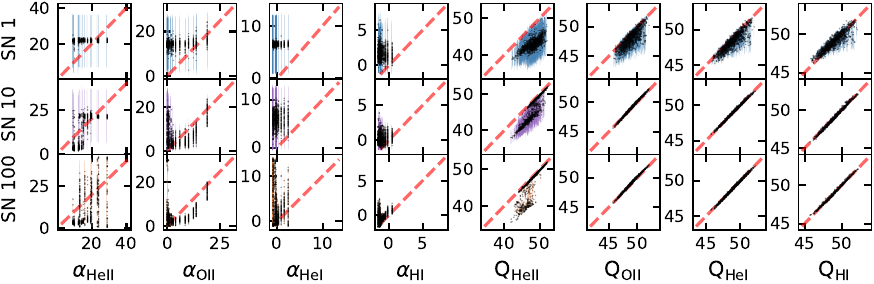}{\textwidth}{(a) ionizing spectrum.}}
    %\hspace*{\fill}
    \gridline{\fig{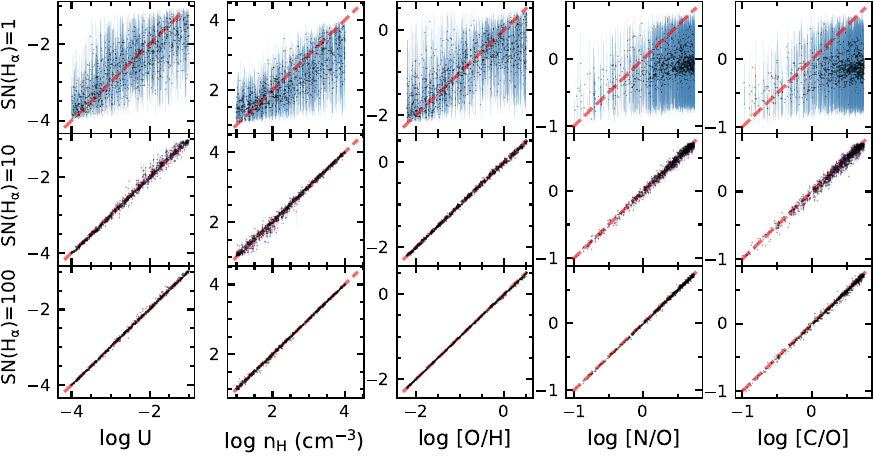}{\textwidth}{(b) nebular properties.}}
    \caption{Mock recovery test for objects with post-AGB type ionizing spectrum. From the top to the bottom row: S/N (H$_\alpha$) $= 1, 10, 100$.}
    \label{fig:mock_pagb}
\end{figure*}

\begin{figure*}
    \centering
    \gridline{\fig{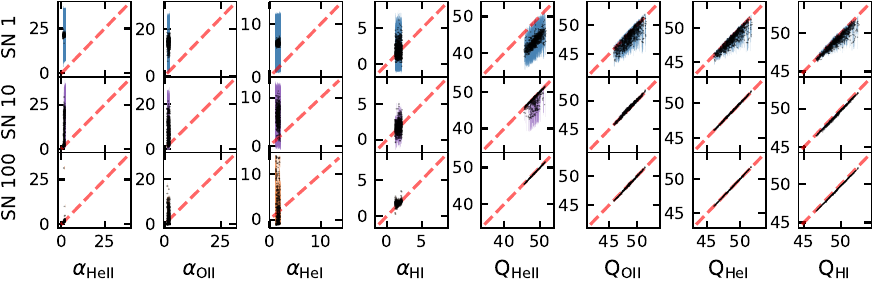}{\textwidth}{(a) ionizing spectrum.}}
    %\hspace*{\fill}
    \gridline{\fig{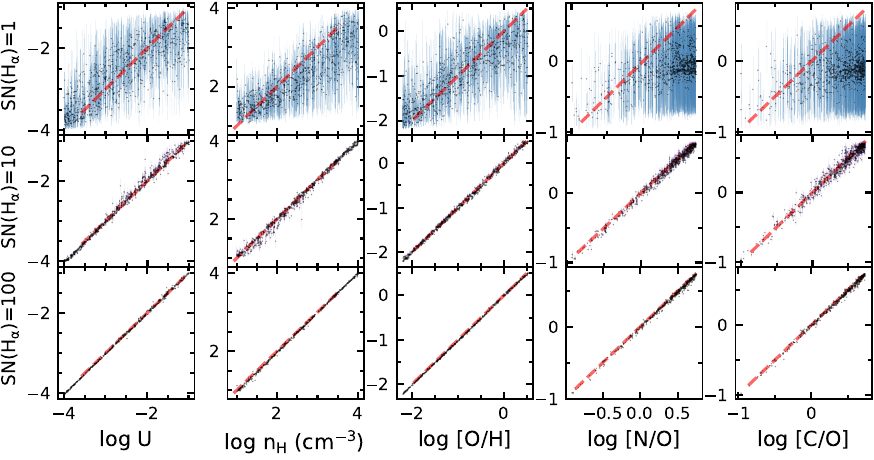}{\textwidth}{(b) nebular properties.}}
    \caption{Mock recovery test for objects with AGN type ionizing spectrum. From the top to the bottom row: S/N (H$_\alpha$) $= 1, 10, 100$.}
    \label{fig:mock_agn}
\end{figure*}

\end{appendix}

\bibliography{main}{}
\bibliographystyle{aasjournal}

\end{CJK*}
\end{document}